\documentclass{article}
\usepackage{cite}
\usepackage[font=small]{caption}

\usepackage{import}
\usepackage{arxiv_style}
\usepackage[centering,hmargin=2.2cm,vmargin=2.5cm]{geometry}
\usepackage[sectionbib]{chapterbib} 
\usepackage{multicol}
\usepackage{authblk}
\numberwithin{equation}{section}

\title{Ehrenfest's theorem beyond the Ehrenfest time}
\author[1,2,3]{Felipe Hernández\footnote{felipeh@mit.edu}}
\author[2,4]{Daniel Ranard\footnote{dranard@mit.edu}}
\author[3]{C. Jess Riedel\footnote{jessriedel@gmail.com}}
\affil[1]{Department of Mathematics, Stanford University, Stanford, CA 94305 USA}
\affil[2]{Center for Theoretical Physics, Massachusetts Institute of Technology, Cambridge, MA 02139, USA}
\affil[3]{NTT Research, Inc., Physics \& Informatics Laboratories, Sunnyvale, CA 94085, USA}
\affil[4]{Department of Physics, California Institute of Technology, Pasadena, CA 91125, U.S.A.}
\date{}                     
\begin{document}


\let\cleardoublepage\relax
\let\clearpage\relax
\maketitle
\begin{abstract}
In closed quantum systems, wavepackets can spread exponentially in time due to chaos, forming long-range superpositions in just seconds for ordinary macroscopic systems. 
A weakly coupled environment is conjectured to decohere the system and restore the quantum-classical correspondence while necessarily introducing diffusive noise---but at what coupling strength, and under which conditions?  
For Markovian open systems with Hamiltonians of the form $\HQ = \PQ^2/2m+V(\XQ)$ and Hermitian linear Lindblad operators, we prove the quantum and classical evolutions are close whenever the strength of the environment-induced diffusion satisfies $\D \gg (\hbar/s_H)^{4/3} \D_H$, where $s_H$ and $\D_H$ are characteristic action and diffusion scales that we define precisely using the classical Hamiltonian $H$.
The bound applies for all observables and for times exponentially longer than the Ehrenfest timescale, which is when the correspondence can break down in closed systems.  
The strength of the diffusive noise can vanish in the classical limit to give the appearance of reversible dynamics. The $4/3$ exponent may be optimal, suggested by heuristic arguments and prior numerical evidence.  
Based on our bound, we give an efficient classical algorithm for simulating quantum Lindblad dynamics, which becomes provably accurate when the strength of environmental coupling exceeds the above threshold. 
\end{abstract}
\vspace{0.5cm}

\begin{figure*}[t!]
\centering
\includegraphics[width=\linewidth, valign=t,scale=0.99]{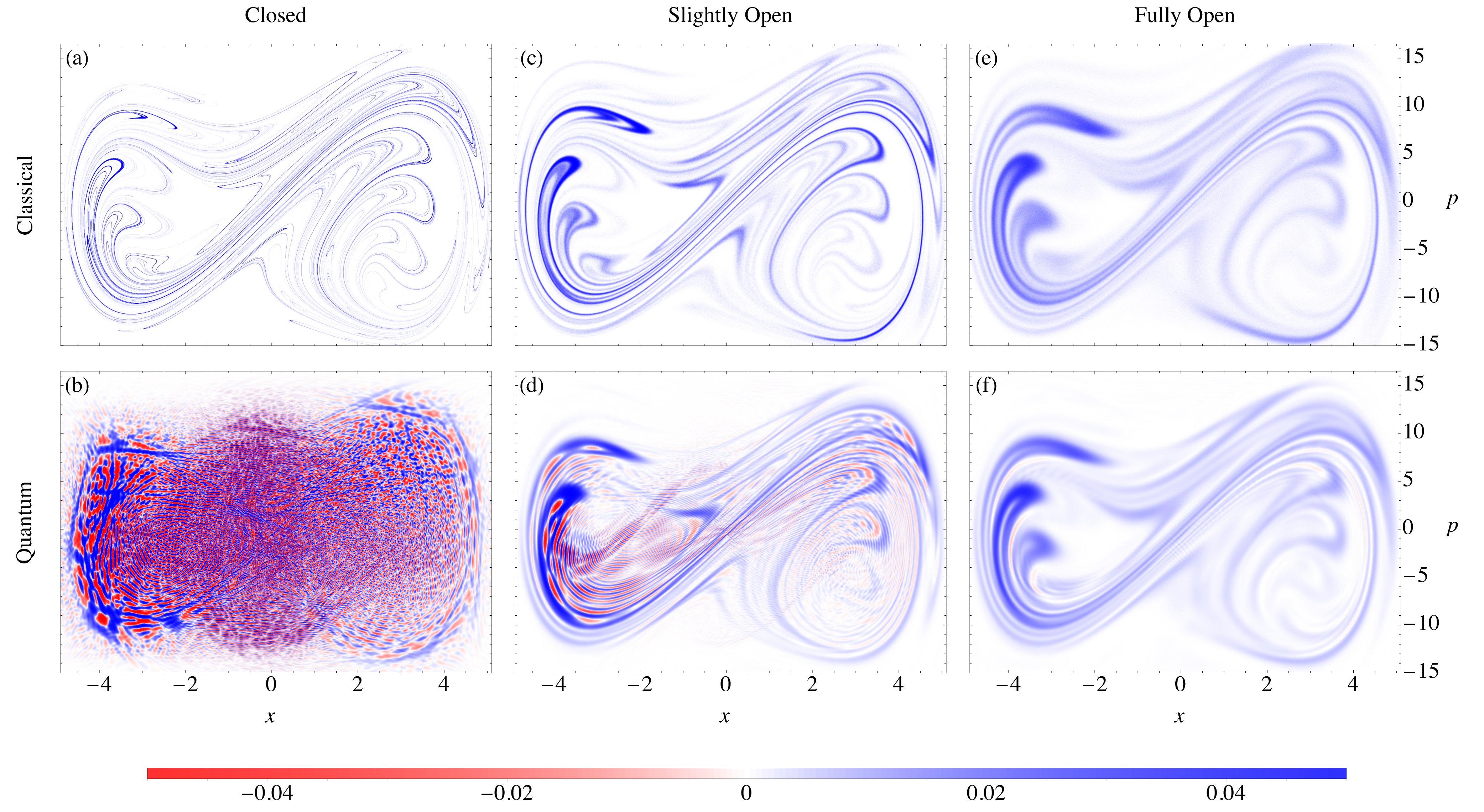}
\caption{
We compare quantum and classical evolutions, for both open and closed  systems, using the example of a chaotic, sinusoidally-driven double-well quartic oscillator. We confirm that when the open system has sufficient diffusion (induced by the environmental coupling), the quantum and classical evolutions agree. While the figure illustrates a special case, Theorem~\ref{thm:mainResult} identifies the sufficient diffusion strength $\D$ at which this transition happens in general systems. The driven oscillator uses Hamiltonian $\HC(\XC,\PC,t) = \PC^2/2m + B \XC^4 - A\XC^2 + C \XC \cos( \omega t)  $ with $A=C=10$, $B=1/2$, $m=1$, $\omega = 6.07$, evolved from Gaussian initial condition $\alpha_0 = (x_0,p_0)=(-3,8)$, $\sigma = (\sigma^{\x\x},\sigma^{\p\p})=(1/10,1)$ for time  $T=16\pi/\omega\approx 8.3$. 
(Cf. Ref.~\cite{habib1998decoherence}.) 
The quantum system uses $\hbar=1/5$ and is simulated with the QuTiP framework \cite{johansson2013qutip}. 
The ``fully open'' versions additionally have homogeneous diffusion matrix $\D \approx \diag(0.0010, 0.13)$, and the ``slightly open'' versions have weaker diffusion $\D/10$;  in the quantum cases these correspond to Lindblad operators proportional to $\XC$ and $\PC$.
(a) The probability distribution of the closed classical system is strictly positive (blue) and develops features on arbitrarily small scales for large times.
(b) In contrast, the Wigner function for the closed quantum system exhibits dramatic non-classicality, as witnessed by strong rapidly oscillating negativity (red) that indicates long-range coherence over phase space. Over time, the smallest features approach minimum scale $(\delta x,\delta p) \sim (\hbar/R_\p,\hbar/R_\x) \sim (0.007,0.02)$, where $(R_\x,R_\p)\sim(10,30)$ are the dimensions of the energetically accessible region of phase space \cite{zurek2001subplanck}.  The vertical bulge in the classically forbidden region near $(x,p)\sim (0,10)$ reflects long-range coherence between $(\pm 3,10)$; the chance of a Gaussian POVM actually measuring the particle near $(0,10)$ is very small.
(c) The slightly open classical system is softened by noise, with less delicate structure than the closed case. (d)  The slightly open quantum system has substantially suppressed long-range coherence, as seen from the elimination of the highest frequency modes (e.g., the vertical bulge at $(0,10)$ is gone), but still exhibits shorter-range coherence and Wigner negativity.
Finally, the fully open classical (e) and quantum (f) systems are highly similar and nearly indistinguishable for any observable. In particular the state of the open quantum system is approximately a mixture of Gaussian wavepackets, necessarily producing an approximately positive Wigner function.}
\label{fig:wigner-simulation}
\end{figure*}

\begin{multicols}{2}
\setcounter{tocdepth}{1} 
\phantomsection
\addcontentsline{toc}{section}{Introduction}

We study the macroscopic emergence of classical mechanics from quantum mechanics in the limit when $\hbar$ is small compared to the characteristic features of the system. This is a well-studied topic in the case of \emph{closed} quantum systems -- that is to say, in Hamiltonian systems isolated from any influence from an external environment -- where classical and quantum observables are known \cite{egorov1969canonical, hagedorn2000exponentially, silvestrov2002ehrenfest} 
by Egorov's theorem \cite{egorov1969canonical} to match closely for times up to the Ehrenfest time $\teh \sim \lambdaL^{-1} \log(s_H/\hbar)$.  This timescale is governed by the dominant Lyapunov exponent $\lambdaL$ and the characteristic action scale $s_H$ of the classical dynamics, and it quantifies the time for a minimal uncertainty wavepacket to spread significantly due to chaos. The above correspondence arises essentially because of Ehrenfest's theorem \cite{ehrenfest1927bemerkung}, which implies that a localized
wavepacket will approximately follow a classical equation of motion as long as it remains well-localized.

For real macroscopic systems, the Ehrenfest time can be quite short---even seconds or minutes---because the dependence on $\hbar^{-1}$ is only logarithmic \cite{berman1978condition, berry1979quantum, chirikov1988quantum}.    
In closed systems beyond the Ehrenfest time, even as $\hbar \to 0$, the correspondence between the classical and quantum evolution breaks down. In particular,
\begin{enumerate}[(1)]
    \item superpositions over macroscopic distances are generated, detectable through delicate interference experiments \cite{haake1987classical, zurek1994decoherence, zurek1998decoherence, habib1998decoherence, berry2001chaos, silvestrov2002ehrenfest} (cf.~\cite{casati1995comment, wiebe2005quantum, schlosshauer2008classicality}); 
    relatedly, the Wigner function develops negativity \cite{berry1979quantum, korsch1981evolution,zurek2001subplanck} (cf.~\cite{tomsovic1991semiclassical}); 
    see Fig.~\ref{fig:wigner-simulation}.
    \item large differences possibly arise between quantum expectation values and the corresponding classical predictions even for smooth observables like $\XQ^2$, according to some numerical studies ~\cite{karkuszewski2002breakdown, carvalho2004environmental} (cf.~\cite{emerson2002quantum}).
\end{enumerate}
In other words, the $\hbar \to 0$ and $t \to \infty$ limits do not commute in closed systems: if one fixes a time duration $t>0$ and takes $\hbar \to 0$, the quantum state trajectory approaches the classical state trajectory, but if one fixes arbitrarily small $\hbar > 0$ and takes $t\to\infty$, then the quantum trajectory may develop superpositions over macroscopic distances. The limit is \emph{singular} in the sense of Berry \cite{berry1995asymptotics,berry2001chaos}.

Despite this theoretical breakdown, 
macroscopic systems appear to obey the laws of classical mechanics for much longer times. To theoretically justify a quantum-classical correspondence beyond the Ehrenfest time, one longstanding suggestion is to consider decoherence effects from the environment, acknowledging that macroscopic systems are rarely well-isolated \cite{zeh1973quantum, zurek1981pointer, zurek1994decoherence, shiokawa1995decoherence, zurek1998decoherence} 
(cf.~\cite{ballentine1994inadequacy,fox1994chaos1,schlautmann1995measurement,casati1995comment,emerson2002quantum,wiebe2005quantum, kofler2007classical}).
Numerical simulations and analytical arguments suggest that decoherence successfully restores the quantitative agreement between the quantum and classical evolution of many specific systems \cite{spiller1994emergence, habib1998decoherence, bhattacharya2000continuous, bhattacharya2003continuous, toscano2005decoherence}.
In particular, it was conjectured that there is a regime where the system-bath coupling is (i) large enough for decoherence to ensure classicality by inhibiting long-range coherence in phase space,
but still (ii) small enough that the noise introduced by the bath does not significantly alter the classical dynamics on fixed time scales, so the system \emph{appears} well-isolated \cite{bhattacharya2000continuous, bhattacharya2003continuous, toscano2005decoherence}.

In this work we prove several aspects of the quantum-classical correspondence suggested above using decoherence modeled by the Lindblad equation.  Our basic strategy is to generalize the Glauber-Sudarshan P function \cite{glauber1963coherent, sudarshan1963equivalence} 
to include \emph{variously} but \emph{boundedly} squeezed Gaussian pure states and then evolve these states with the local harmonic approximation to the dynamics (i.e., an open-system generalization of Heller's thawed-Gaussian approximation 
\cite{heller1975time}).

The results presented here apply to Hamiltonians of the common form $\HQ=\PQ^2/2m + V(\XQ)$ with Hermitian linear Lindblad operators (and hence no friction). 
(The case of general Hamiltonians $\HQ(\XQ,\PQ)$ and Lindblad operators $\LQk(\XQ,\PQ)$ is presented in companion work \cite{hernandez2023decoherence2}.)
Such a system may represent multiple particles in multiple spatial dimensions, with general position-dependent interactions. (The kinetic term could be replaced with a general quadratic function of momenta, allowing particles with distinct masses, but we omit this variation for simplicity.)  With $N$ particles in $n$ spatial dimensions, we have $d=Nn$ degrees of freedom.

The Hermitian linear Lindblad operators used here may be seen as the first-order approximation of a general environmental interaction, and their effect is to introduce both decoherence and noise in phase space.  
The momentum variance $\langle \hat{p}^2\rangle$ of a Gaussian wavepacket  grows like the momentum diffusion rate, $\dot\sigma^{\p\p}\sim\Dp$, while the coherence between wavepackets separated by distance $\Delta x$ decays at the ``localization rate'' $\Lambdax\Delta x^2$ for $\Lambdax = \hbar^{-2}\Dp$ \cite{joos1985emergence,zurek1991decoherence}.  (The same holds under exchange $x\leftrightarrow p$.) Because of the factor of $\hbar^{-2}$, we can simultaneously have $\Lambdax\to\infty$ and $\Dp\to 0$ in the classical limit $\hbar\to 0$, giving the appearance of reversible noiseless dynamics while keeping macroscopic superpositions suppressed.

We prove that given sufficient diffusion, for times $t \ll \hbar^{-1/2},$
the quantum evolution under the Lindblad equation is well-approximated by the above classical evolution for all possible observables, in a sense we will make precise.  
This timescale is exponentially longer than the Ehrenfest time (as defined via local Lyapunov exponent; see Section \ref{sec:main-result}).
Indeed, for a macroscopic system with characteristic action scale 
$s_H\sim 1\,\mathrm{kg}\cdot\mathrm{m}^2/\mathrm{s}$ and Lyapunov exponent $\lambdaL \sim 1\,\mathrm{s}^{-1}$, the Ehrenfest timescale is $\teh \sim \lambdaL^{-1} \log(s_H/\hbar) \sim 1$ minute, while the bound presented below extends to the timescale $\tau\sim \lambdaL^{-1} \sqrt{s_H/\hbar} \sim 10^{17}\, \mathrm{s} \sim 1$ billion years.

We find the above quantum-classical correspondence holds when the environment-induced diffusion satisfies $D \gg \hbar^{4/3}$. Remarkably, evidence from Toscano et al.\ 
\cite{toscano2005decoherence,wisniacki2009scaling} 
suggests this condition is actually necessary, in contrast to the weaker condition $D \gg \hbar^2$ often suggested \cite{kolovsky1994remark, zurek2003decoherence, pattanayak2003parameter};
then $D \sim \hbar^{4/3}$ may be a genuine threshold for the correspondence.  We  offer a heuristic argument later with Eq.~\eqref{eq:moyal}.

\section{Preliminaries}\label{sec:prelim}

For a quantum system of finite $d$ degrees of freedom, 
we consider the quantum evolution $\rho(t)$ under the Lindblad equation $\partial_t \rho(t) = \LLQ [\rho(t)]$ with Lindbladian
\begin{align} \label{eq:lindblad-simple}
	\LLQ [\rho] &= \frac{-i}{\hbar}[\HQ,\rho]+\frac{1}{\hbar}\sum_{k} (\LQk \rho \LQk^\dagger-\frac{1}{2}\{\LQk^\dagger\LQk,\rho\}).
\end{align}
We treat the case of Hamiltonian $\HQ=\PQ^2/2m + V(\XQ)$ with Hermitian linear Lindblad operators $\LQ{k}=\ell_{k}\cdot(\XQ,\PQ)=\ell_{k,\x}\cdot\XQ+\ell_{k,\p}\cdot\PQ$ for $\ell_{k} = (\ell_{k,\x}, \ell_{k,\p})\in\mathbb{R}^d\times\mathbb{R}^d$.
The corresponding classical dynamics are given by a frictionless Fokker-Planck equation $\partial_t \cstate = \LLC[\cstate]$ using the Liouvillian 
\cite{risken1984fokkerplanck}
\begin{equation}
	\label{eq:fp-simple}
	\LLC[\cstate] = - \sf^{ab}(\partial_a \cstate)(\partial_b \HC)  +	\frac{1}{2} \D^{ab} \partial_a \partial_b \cstate,
\end{equation}
where the first term above produces classical Hamiltonian flow using $\HC=\PC^2/2m + V(\XC)$, while the second term above produces diffusion described by the positive semidefinite matrix $\D^{ab} = \hbar \sf^{ac}  \sf^{bd}\sum_k \Real \ell_{k,a}\ell_{k,b}^*$. Here, $\sf = \big(\begin{smallmatrix}0 & \IdM_d \\-\IdM_d & 0 \end{smallmatrix}\big)$ is the (antisymmetric) symplectic form and $\partial_a$ denotes the partial derivative in phase space, where 
the indices $a,b \in\{ \x_1,\ldots,\x_d,\p_1,\ldots,\p_d\}$ range over the $2d$ directions in phase space and
repeated indices are summed.

Just as the diffusion equation arises from a random walk, the Fokker-Planck equation in~\eqref{eq:fp-simple} could also be interpreted as arising from a Langevin equation \cite{kerr2000generalized}, describing noisy trajectories on phase space. Thus~\eqref{eq:fp-simple} describes an ensemble of noisy classical trajectories, each roughly following the classical equation of motion when $D$ is small.

We identify some key scales in any Hamiltonian $\HQ = \PQ^2/2m + V(\XQ)$. At each position $x$ there is a local harmonic approximation to the dynamics depending on the Hessian $\nabla^2 V(x)$ of the potential. 
The \introduce{harmonic time} $\tH := \sqrt{m/\JkSN{V}{2}} \lesssim \lambdaL^{-1}$ is the shortest timescale associated with such local dynamics, 
where 
\begin{align}
	\JkSN{V}{j} := 
	\sup_{x} \sup_{\|w_i\|=1} \left| \left[\prod_{i=1}^j (w_i \cdot \nabla)\right] V(x)\right|
\end{align}
is the maximum over space of $j$ directional derivatives of the potential. Similarly, the \introduce{harmonic aspect parameter} $\unitRatio_H :=  \sqrt{m\JkSN{V}{2}} = m/\tH = \JkSN{V}{2} \tH$ (with units of [momentum/length] $\sim$ kg/s) picks out a preferred ``aspect ratio'' in phase space. The \introduce{anharmonic action} $s_H :=  \unitRatio_H^3/(\tH \JkSN{V}{3})^{2} = m^{1/2}\JkSN{V}{2}^{5/2}\JkSN{V}{3}^{-2}$ is the action scale above which anharmonicities of the potential (quantified by the max \emph{third} derivative $\JkSN{V}{3}$) are important. 
Finally, 
\begin{align*}
	\D_H := \frac{s_H}{\tH} \left(\begin{matrix}  \unitRatio^{-1}_H\IdM_d & 0\\ 0  &  \unitRatio_H\IdM_d  \end{matrix}\right)
\end{align*}
is the \introduce{anharmonic diffusion matrix}. In making use of the above scales, we assume $\JkSN{V}{2}, \JkSN{V}{3}$ are finite.

Let $\tauQas$ and $\tauCas$ denote the normalized Gaussian quantum state (a density matrix) and classical state (a distribution) 
with mean $\alpha = (\alpha^\x,\alpha^\p)\in\bbR^{2d}$ in phase
space and with covariance matrix $\sigma \in\bbR^{2d} \times \bbR^{2d}$.
The above parameters naturally define a preferred covariance matrix 
\begin{align*}
	\sigmaco:=  \frac{\hbar}{2} \left(\begin{matrix}  \unitRatio^{-1}_H\IdM_d & 0\\ 0  &  \unitRatio_H\IdM_d  \end{matrix}\right) = \frac{\hbar \tH}{2 s_H} \D_H 
\end{align*}
and the corresponding (unsqueezed) \introduce{coherent states} $\tauQ_{\alpha,\sigmaco}$, which are pure state that saturate the uncertainty principle. 
Our result utilizes a special subset of all normalized pure Gaussian quantum states that we call \introduce{not too squeezed} (NTS): 
the states are squeezed relative to the coherent states by no more than a factor
$\zz := \max\{\frac{\hbar/s_H}{\Dz},1\}  \ge 1$,
where the minimum eigenvalue 
\begin{align}\label{eq:D0}
	\Dz := \lambdamin\left(\D_H^{-1/2}\D \D_H^{-1/2}\right)
\end{align}
is a dimensionless measure of the diffusion strength relative to $\D_H$. In other words, NTS states $\tauQas$ are pure and satisfy
$\sigma \le \sigmaNTS$ where
\begin{align}\label{eq:NTSsigma}
	\sigmaNTS :=\zz\, \sigmaco = \frac{\hbar}{2} \max\left\{\frac{\hbar/s_H}{\Dz},1\right\}  \left(\begin{matrix}  \unitRatio^{-1}_H\IdM_d & 0\\ 0  &  \unitRatio_H\IdM_d  \end{matrix}\right).
\end{align}

\begin{figure*}[!b]
\centering
\includegraphics[width=\linewidth, valign=t,scale=0.8]{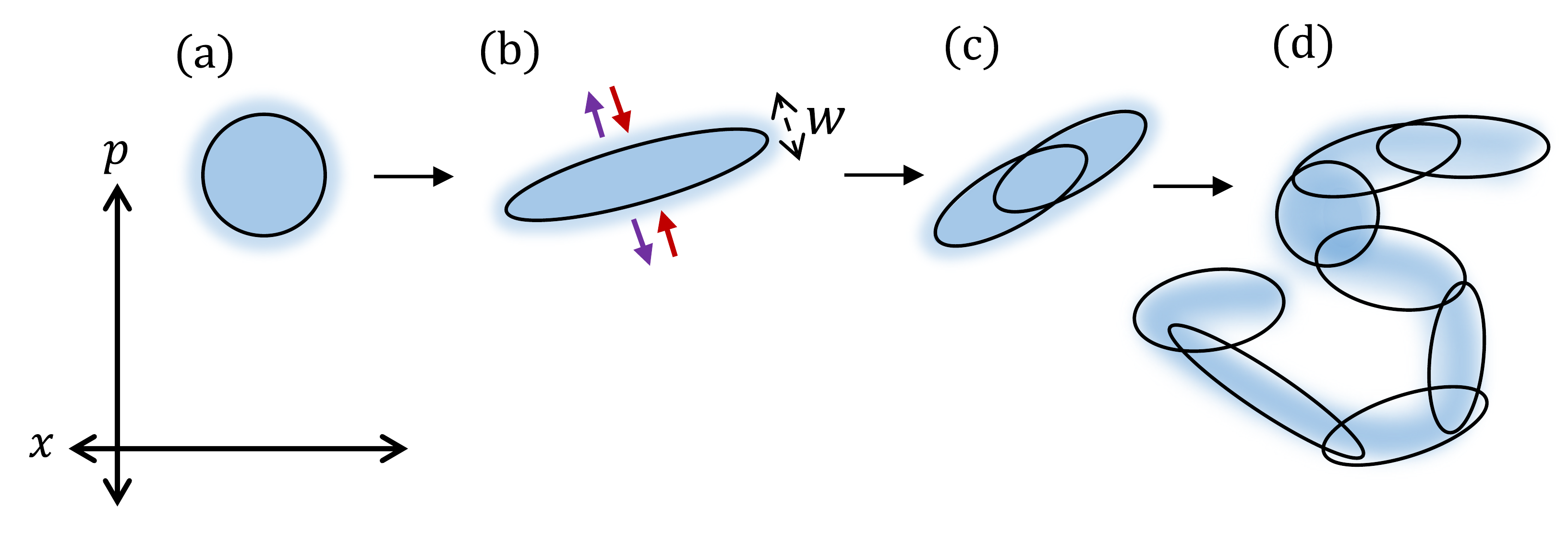}
\caption{
	(a) The Wigner function for an initially pure quantum Gaussian state  $\rho(t \liq 0)$ evolves in phase space.
	(b) At short times the dynamics admit a local harmonic (quadratic) approximation, broadening the distribution via diffusion (purple arrows) and possibly squeezing it via classical flow (red arrows). For diffusion strength $D$ and local Lyapunov exponent $\lambdaL$ of the flow, the Gaussian state (ellipse) has a minimum thickness: the diffusion broadens the ellipse at speed $\dot{w} \sim D/w$, while the the Hamiltonian flow can shrink the width by at most $\dot{w} \sim -w /\lambdaL$, with the competing effects balanced at $w \sim (D/\lambdaL)^{1/2}$. 
	(c) After $\rho(t)$ becomes mixed due to diffusive broadening, it can be approximated by a mixture $\rhot(t)$ of pure Gaussian states (ellipses) that are individually less squeezed. Each evolves by its own local harmonic dynamics while continuously being further decomposed.
	(d)  As $\rho(t)$ spreads in phase space, our approximation $\rhot(t)$ uses ellipses of fixed area $\hbar$ but varying amounts of squeezing. See Eq.~\eqref{eq:simple-trajectories}.}
\label{fig:ellipses}
\end{figure*}

We will assume a quantum system is initially a (possibly trivial) mixture of NTS quantum states, i.e., 
\begin{align}
\label{eq:simple-initial-quantum-state}
\rho(t \liq 0):=& \intNTS  \, \pas(t \liq 0) \tauQas
\end{align}
where $\pas(t \liq 0)\ge $ is a probability distribution supported only on 
$\sigma \le \sigmaNTS$. 
We will compare it to the analogously initialized classical system,
\begin{align}
\label{eq:simple-initial-classical-state}
f(t \liq 0):=& \intNTS \, \pas(t \liq 0) \tauCas,
\end{align}
Our main result states that, given sufficient diffusion, the corresponding quantum and classical evolutions will continue to be well-approximated by mixtures of NTS states with a single underlying distribution $\pas(t)$ at all time. 

\begin{figure}[H]
\includegraphics[width=\linewidth, valign=t,scale=1]{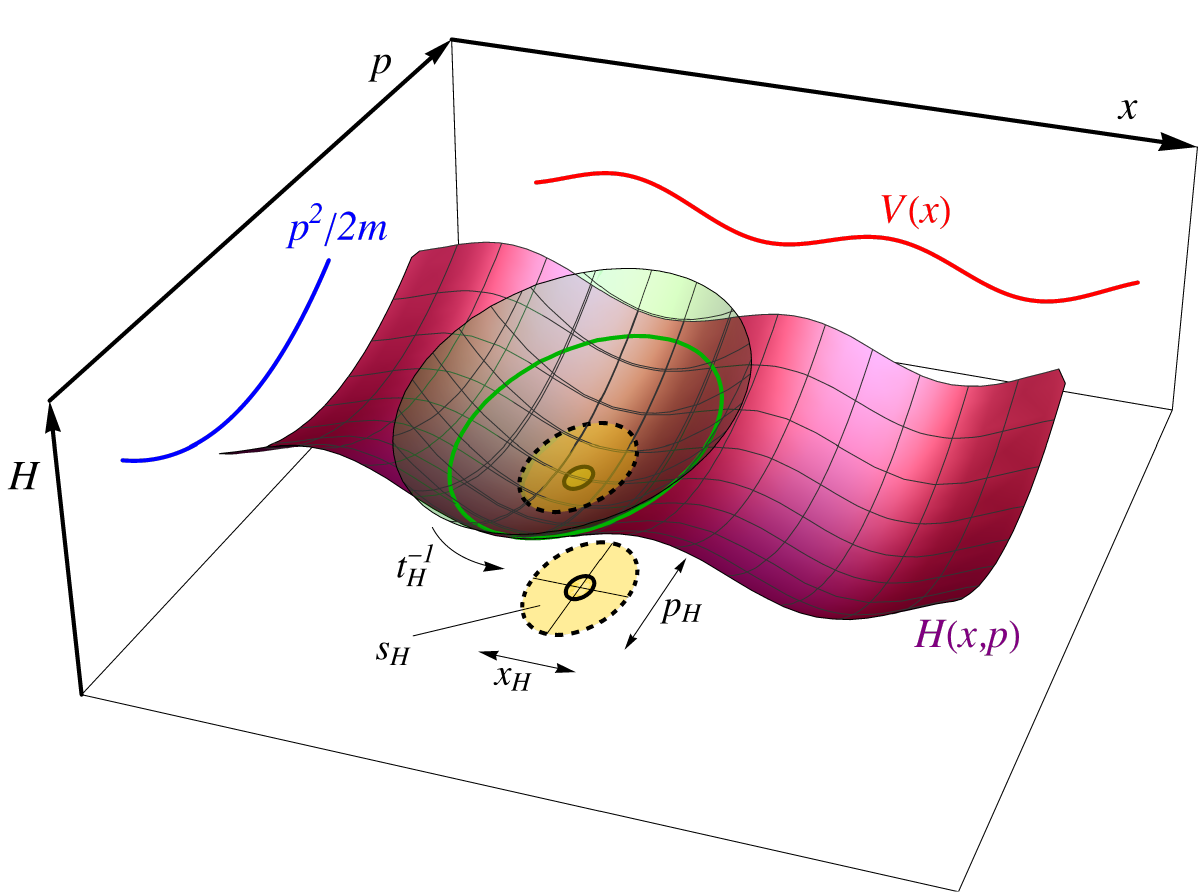}
\caption{
    An illustration of the characteristic scales of a Hamiltonian $\HC = \PC^2/2m + V(\XC)$, which defines the opaque purple surface over phase space. The potential $V(\XC)$ is projected in red and the kinetic term $\PC^2/2m$ is in blue. For this potential, the second derivative $\nabla^2 V$ takes its largest absolute value near a local minimum. The transparent green paraboloid denotes the second-order approximation to the Hamiltonian (i.e., the conservative part of the local harmonic dynamics) near this point. The inverse harmonic time $\tH^{-1} = \sqrt{\JkSN{V}{2}/m}$ is the characteristic frequency, at any amplitude, of the harmonic motion within this local approximation to the potential (e.g., around the thick green ellipse). The small solid-black ellipse represents a classical (Gaussian) Gibbs state for this quadratic potential.   Although the area of the small ellipse is temperature dependent, the harmonic aspect parameter $\unitRatio_H = \sqrt{m \JkSN{V}{2}}$ of the ellipse --- the ratio of its axes with units kg/s --- is not. This determines a natural choice of (unsqueezed) quantum coherent states, corresponding to similarly oriented ellipses in phase space with area $\sim \hbar$.  The harmonic approximation will be accurate within the anharmonic length $x_H = \sqrt{s_H /\unitRatio_H} = \JkSN{V}{2}/\JkSN{V}{3}$ in the $\XC$ direction and the anharmonic momentum $p_H =  \sqrt{s_H \vphantom{/}\unitRatio_H}  = \sqrt{m \JkSN{V}{2}^3}/\JkSN{V}{3}$ in the $\PC$ direction, which are determined by the largest anharmonicity $\JkSN{V}{3}$ and which satisfy $\unitRatio_H = p_H/x_H$. This characterizes the region of harmonic approximation accuracy, the dotted-black ellipse, which has phase-space area of order the anharmonic action $s_H = x_H p_H = m^{1/2}\JkSN{V}{2}^{5/2}/\JkSN{V}{3}^{2}$, the unique action scale constructible from $m$, $\JkSN{V}{2}$, and $\JkSN{V}{3}$.  
    The anharmonic diffusion matrix $\D_H =  2s_H\sigmaco/(\hbar\tH) =\mathrm{diag}(x_H^2,p_H^2)/\tH$ is a characteristic diffusion scale above which diffusion appreciably affects the system, relative to the largest anharmonicities, on the harmonic time scale. As quantified in Theorem~\ref{thm:mainResult} and Corollary~\ref{cor:Main}, quantum and classical dynamics will be essentially indistinguishable for \emph{all} observables when $\D \gg (\hbar/s_H)^{4/3} \D_H$, where $\D$ is the actual diffusive noise from the open-system dynamics, and where $\hbar/s_H \ll 1$ for macroscopic systems.
 }
\label{fig:params}
\end{figure}

\section{Main result and implications}\label{sec:main-result}

\phantomsection
\addcontentsline{toc}{section}{Main result}

We can now state our main result.

\begin{restatable}[Main result]
{thm}{thmHamVUnitsOLD}\label{thm:mainResult}
Suppose $\rho(t)$ solves the Lindblad equation \eqref{eq:lindblad-simple} with initial condition satisfying  \eqref{eq:simple-initial-quantum-state}, e.g., $\rho(0) = \tauQ_{\sigma_*,\alpha_0}$, and let $\cstate(t)$ be the unique classical trajectory solving the corresponding Fokker-Planck equation \eqref{eq:lindblad-simple} with initial condition \eqref{eq:simple-initial-classical-state}.
Then there exists a time-dependent extension $\pas(t)\ge 0$ of the initial probability distribution for all $t\ge 0$ that defines both of the approximating trajectories
\begin{align}\label{eq:simple-trajectories}
\begin{split}
	\rhot(t):=& \intNTS \, \pas(t) \tauQas\\
	\tilde{f}(t):=& \intNTS  \, \pas(t) \tauCas
\end{split}
\end{align}
such that
\begin{align}\label{eq:bound-simple-potential}
	\begin{split}
		\|\,\rhot(t) - \rho(t)\|_{\mathrm{Tr}\,} &\leq  \er t\\
		\|\tilde{f}(t) - f(t)\|_{L^1} & \leq \er t
	\end{split}
\end{align}
with error rate
\begin{align}\label{eq:bound-simple-potential-eps}
	\er
	&= d^{\frac32}\frac{1}{\tH} \sqrt{\frac{\hbar}{s_H}} \max\left\{\frac{\hbar/s_H}{\Dz}, 1\right\}^{\frac32}.
\end{align}
\end{restatable}
\noindent Note that the classical approximate trajectory $\cstatet$ is simply the \introduce{Wigner function} of the quantum approximate trajectory $\rhot$, that is $\tilde{f}(t)= \WW[\rhot(t)]$, because $\tauCas = \WW[\tauQas]$ (see Methods Section~\ref{sec:wigner}).

A proof of Theorem~\ref{thm:mainResult} is sketched in Section~\ref{sec:sketch} and given in detail in the Methods Section~\ref{sec:proof}. We consider classical simulation, asymptotic optimality, and physical examples in Sections~\ref{sec:classical-simulation-claim}, \ref{sec:optimality}, and \ref{sec:physical-examples}.  Before turning to these, let us interpret the result.

Recall that $d$ is the number of degrees of freedom, and that $\tH$, $s_H$, and $\D_H$ above are just characteristic scales defined with the parameters $m$, $\JkSN{V}{2}$, and $\JkSN{V}{3}$ of the classical Hamiltonian $\HC=\PC^2/2m+V(\XC)$.  
In \eqref{eq:bound-simple-potential}, $\|\hat{A}\|_{\mathrm{Tr}}:=\Tr[(\hat{A}^\dagger \hat{A})^{1/2}]$ is the trace norm on quantum operators and $\|A\|_{L^1} := \int\! \dd \alpha |A(\alpha)|$ is the analogous $L^1$ norm on classical phase-space functions. As recalled in Methods Section~\ref{app:norm-interp}, they constrain the probability of a discrepancy being revealed by the measurement of any quantum observable $\hat{A}$ or classical observable (function on phase space) $A$. 
We thus have the following corollary.
\begin{restatable}[Indistinguishable observables]
{cor}{corMain}\label{cor:Main}
Under the conditions of Theorem~\ref{thm:mainResult}, the expectation values of any quantum observable $\hat{A}$ and its Wigner-Weyl-transformed classical counterpart $A$ (a function on phase space) obey 
\begin{align}\nonumber
	\left|\Tr[\hat{A}\rho(t)] - \int \! A(\alpha)f(t,\alpha)\mathrm{d}\alpha\right| \leq  
    (\|\hat{A}\|_{\mathrm{op}}+\|A\|_{L^\infty})  \er t 
\end{align}
where $\er$ is given by \eqref{eq:bound-simple-potential-eps}.
\end{restatable}
\noindent 
Here, $\|\cdot\|_{\mathrm{op}}$ and $\|\cdot\|_{L^\infty}$ denote the operator norm and the essential supremum norm.
This corollary follows from the Weyl trace formula $\Tr[\hat{A}\rhot(t)] = \int \! A(\alpha)\WW[\rhot](t)(\alpha)\mathrm{d}\alpha$.

How much environment-induced diffusion is necessary for Theorem \ref{thm:mainResult} to  ensure a close quantum-classical correspondence?  First note from \eqref{eq:bound-simple-potential-eps} that regardless of the strength of diffusion, our construction can never  bound the \emph{rate} at which the 
quantum-classical error accumulates to be lower than
\begin{equation} \label{eq:eps-thresh}
\ermax
:=d^{\frac32} \frac{1}{\tH}\sqrt{\frac{\hbar}{s_H}}.
\end{equation}
Then, given a desired error rate $\er\ge \ermax$, we can see the required diffusion strength is 
\begin{equation}\label{eq:diff-thresh}
\D \geq \left( \frac{\er }{d^{\frac32}\tH^{-1}}\right)^{-\frac23} \left(\frac{\hbar}{s_H}\right)^{\frac43} \D_H .
\end{equation}
See Fig.~\ref{fig:ellipses} 
for a heuristic argument suggesting the adequacy of condition~\eqref{eq:diff-thresh}.

Alternatively, if $\D \gtrsim (\hbar/s_H)^{\frac43-p}\D_H$ for some power $p>0$, we find that the total error $\epsilon=\er t$ in the correspondence~\eqref{eq:bound-simple-potential} is small for times $t \lesssim \tH (s_H/\hbar)^q$, for power $q=\min\{\frac12,\frac{3p}{2}\}$. This time range is exponentially longer than the Ehrenfest time.  Here we refer to the Ehrenfest time as $\teh \sim \lambdaL^{-1} \log(\hbar^{-1})$ where $\lambdaL$ is the maximum \textit{local} or \textit{instantaneous} Lyapunov exponent given by $\lambdaL \sim \tH^{-1}  =\sqrt{\JkSN{V}{2}/m}$, the maximum nonlinearity of the classical flow. 
This quantity upper bounds the \emph{global} Lyapunov exponent, defined by an infinite-time limit.

\begin{figure}[H]
	\centering
	\includegraphics[width=\linewidth, valign=t,scale=0.9]{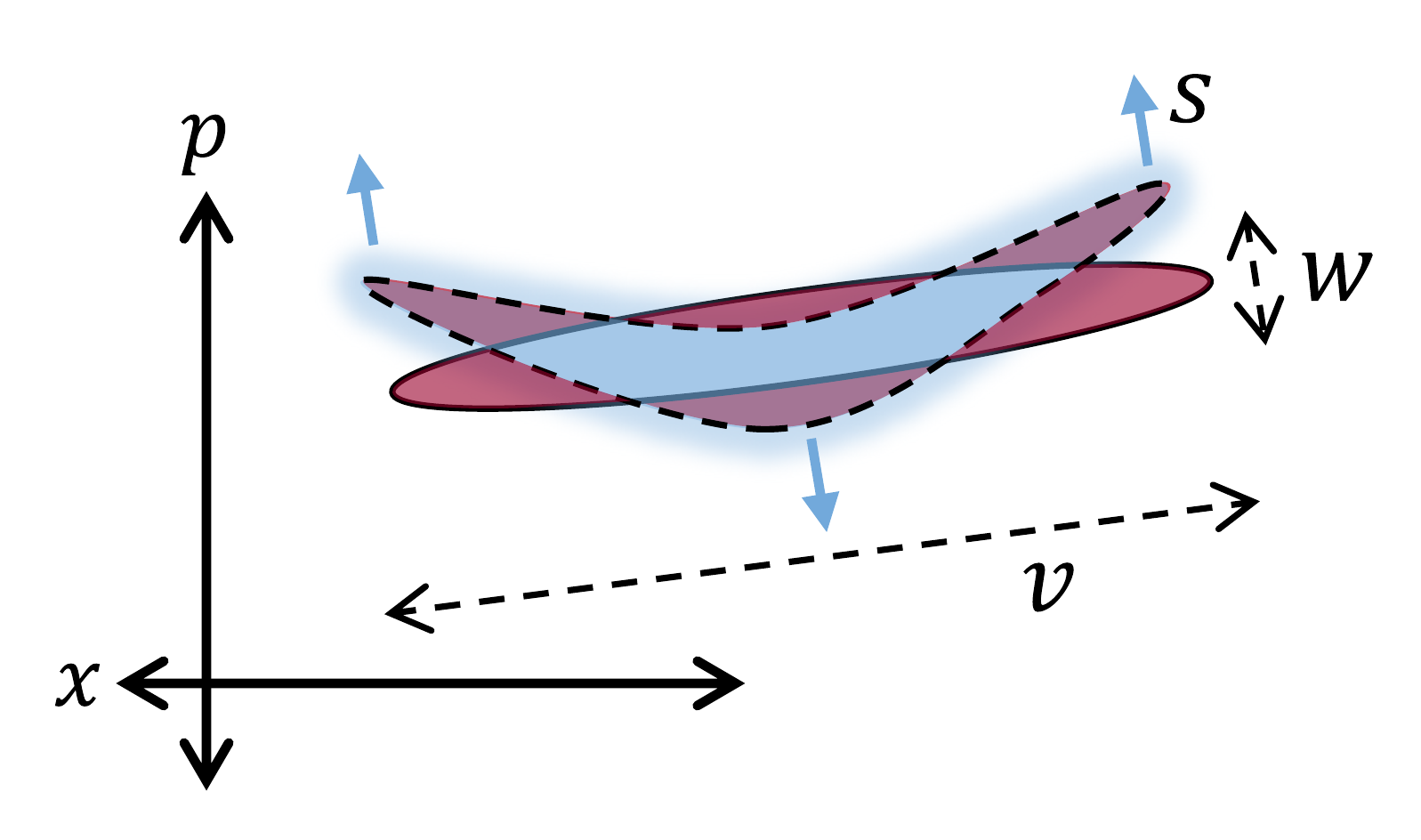}
	\caption{
		Consider the error rate incurred by evolving a pure Gaussian state (ellipse) using the local Harmonic approximation to the dynamics.     
		The discrepancy is dominated by the leading-order anharmonicity $\nabla^3 V$, which is strongest (relative to the center) at the tips of the ellipse lying on either end of the long axis of length $v \sim \hbar/w$.  As the third-order term in the Taylor approximation, this changes the speed of the local flow by $s \lesssim v^2 \|\nabla^3 V\|$. So the discrepancy (red shaded area) between the distribution under the exact dynamics (curved boomerang) and the area of the ellipse grows at rate $\lesssim sv$.  Compared to the ellipse's area $\hbar$, this gives an error rate $\er \sim sv/\hbar \lesssim  \hbar^{-1} v^3 \|\nabla^3 V\| $.
		Because the eigenvalues of the covariance matrix 
		$\sigma \sim \textrm{diag}(w^2,v^2)$ 
		are just the squares of the semimajor axes, the error rate is bounded as $\er \lesssim \hbar^{-1} \|\sigma\|^{\frac32} \|\nabla^3 V\|$, as proved  in Lemma~\ref{lem:HarmErrorQV}.
		As explained in Fig.~\ref{fig:ellipses}, the width $w$ of the narrow axis is lower bounded by the diffusion strength: $w \gtrsim (D/\lambdaL)^{1/2}$.  So we can also write $\er \lesssim  (\hbar^{4/3}/\D)^{3/2} \lambdaL^{3/2} \|\nabla^3 V\|$, which is small when $\D \gg \hbar^{4/3}$.
	}
	\label{fig:harm-err}
\end{figure}

We emphasize that $D$ may be chosen vanishingly small as $\hbar \to 0$, while still yielding small quantum-classical error.  Thus the correspondence holds for long times, even while the system appears isolated during any fixed time interval.  (That is, for any fixed timescale, the effect of the noise on the trajectory vanishes as $\hbar \to 0$.)
Nonetheless, for any fixed nonzero $D$, at sufficiently large time, the effect of $D$ on the classical solution to $\eqref{eq:fp-simple}$ becomes large.  For chaotic evolution, this effect already occurs after time $t \sim \lambdaL^{-1}\log(D^{-1})$, or equivalently the Ehrenfest time if $D \sim \hbar$, because the diffusive noise is exponentially amplified.

\section{Classical algorithm for simulation}
\label{sec:classical-simulation-claim}

Building on Theorem \ref{thm:mainResult}, we offer a classical algorithm to simulate the quantum Lindblad evolution, in the sense of computing observables $\Tr[\hat{A} \rho(t)]$. For $d$ continuous degrees of freedom, a naive classical algorithm to simulate a closed quantum system (say via exact diagonalization, or Schrodinger evolution of the discretized wavefunction) would require time and space complexity $\textrm{poly}(\hbar^{-d})$, due to effective Hilbert space dimension $(S/\hbar)^d$ for a region of phase space of area $S^d$.  However, for an open system with sufficient decoherence, Corollary \ref{cor:Main} enables one to estimate $\Tr[\hat{A} \rho(t)]$ by simulating the classical Fokker-Planck equation rather than the quantum Lindblad equation.  Then the computational complexity becomes
$\textrm{poly}(D_0^{-1}, d)$, 
now depending on the diffusion strength $D_0$ but \textit{not} on $\hbar$, with only $\textit{polynomial}$ dependence on the $d$ degrees of freedom. 

To formalize this result about classical simulation, we consider a slightly modified class of Hamiltonians.  This modification guarantees the system is confined in phase space, avoiding arbitrarily large flows that become problematic for simulation techniques. 
\begin{restatable}[Lindblad simulation (simplified)]{thm}{thmSimulationSimplified}
	\label{thm:simulation-simplified}
    Assume the conditions of Theorem~\ref{thm:mainResult} but with $\HQ = \VP(\PQ)+ V(\XQ)$ where the first three derivatives of $\VP$ and $V$ are bounded.
    Let $T>0$ be a time and $\error \ge 12T \er$ a desired error, with $\er$ given by \eqref{eq:bound-simple-potential-eps} for this Hamiltonian.
    Then there exists a (randomized) classical algorithm to compute the expectation value $\Tr[\rho(t)\hat{A}]$ of any quantum observable $\hat{A}$ satisfying $\|A\|_{L^\infty} + \|\hat{A}\|_{\mathrm{op}}\leq 1$ within error  $\error$ (with probability at least $99\%$) and time-complexity
    \begin{align}\label{eq:sim-complexity-simplified}
        \mathcal{O}\left(\textup{poly}
        (d,\error^{-1},T/\tH,\D_{0}^{-1})\right)
    \end{align}
    independent of $\hbar$.
\end{restatable}
\noindent The constant hidden by $\mathcal{O}(\cdot)$ includes polynomial dependence on $V$ and $\VP$ and their derivatives up to third order and also on $\Dinf := \lambda_{\mathrm{max}} (\D_H^{-1/2}\D\D_H^{-1/2})$, because large forces or diffusion require small time-steps. A stronger version of this theorem is proven in Methods Section~\ref{sec:simulation}, with more explicit parameter dependence than \eqref{eq:sim-complexity-simplified}.  

This result may be seen as a continuous-variable analog to discrete-variable results in noisy random circuit sampling \cite{aharonov2023polynomial}, 
or as an analog to results about noisy boson sampling \cite{qi2020regimes} (but without the restriction to linear dynamics on phase space),
both cases where noiseless closed quantum evolution is infeasible to classically simulate while the noisy version is tractable.

\section{Asymptotic optimality}
\label{sec:optimality}

Theorem~\ref{thm:mainResult} show that the  $D \gtrsim \hbar^{4/3}$ threshold is \textit{sufficient} to guarantee quantum-classical correspondence beyond the Ehrenfest time. We now present a heuristic argument that it is \textit{necessary}, i.e., that our bound is optimal and cannot be asymptotically improved without further assumptions about specific systems.   Denoting the Wigner function as $\W(t)=\WW[\rho(t)]$, it evolves according to the Moyal bracket, expanded as
\begin{equation}
\begin{split}\label{eq:moyal}
\partial_t \W(t) &= \Moyalbracket{H,\W} \\  
& =(\partial_a \HC)(\partial^a \W) - \hbar^2(\partial^3 \HC)(\partial^3 \W) + \ldots
\end{split}
\end{equation}
where $\partial^3$ schematically denotes a sum of third partial derivatives.  The ellipses indicate further terms of the form $(\partial^{2n+1} H) (\partial^{2n+1}\W)$.   If $\W$ has a minimum length $w$ on which it varies (e.g.\ if the distribution has tendrils of minimum width $w$), then $\partial^3 f \lesssim w^{-3} f$, and the leading $\hbar$-dependent term in the above expansion has magnitude $\hbar^2 w^{-3} f$.  Meanwhile, the $\partial^n H$ terms are independent of $\hbar$, so the error between the classical flow of $f$ under $H$ and the corresponding evolution by the Moyal bracket is
\begin{align} \label{eq:wigner-expand-norm}
    \norm{\partial_t f - (\partial_a \HC)(\partial^a f)}_{L_1} \lesssim \hbar^2 w^{-3}.
\end{align}
In a closed chaotic system, the minimum width of variation of $f$ can decrease like $w \sim e^{- \lambdaL t}$ for Lyapunov exponent $\lambdaL$, due to exponential squeezing and stretching, leading to large error within the Ehrenfest time.  Meanwhile, for an open system, the squeezing is balanced by diffusion, leading to a minimum width $w \sim (D/\lambdaL)^{1/2}$ (Fig.~\ref{fig:ellipses}).  Then the error in Eq.~\eqref{eq:wigner-expand-norm} scales as $\hbar^2 D^{-3/2}$, which is large unless $D \gg \hbar^{-4/3}$, suggesting this threshold is necessary for the classical approximation.

\section{Proof sketch}\label{sec:sketch}

The proof requires three steps: defining $\pas(t)$ using the harmonic approximation to the exact dynamics; 
applying the Duhamel formula to bound the harmonic approximation error as a sum (integral) of errors accumulated at each time step; and bounding the error at each time step.


\proofpart{Constructing the distribution $\pas(t)$}
If $\rhot(t)$ is to approximate $\rho(t)$, which satisfies 
the Lindblad equation~\eqref{eq:lindblad-simple},
then we would like to find 
$\pas(t)$ such that 
$\partial_t \rhot(t)\approx \LLQ [\rhot(t)] = \iint\! \dd \alpha \dd \sigma  \, \pas(t) \LLQ [\tauQas]$.
To accomplish this, we will use a \introduce{harmonic approximation} $\LLQLa$ to $\LLQ$ near the point $\alpha$ in phase space. Harmonic dynamics are characterized by a quadratic Hamiltonians and linear Lindblad operators, and they exactly preserve the set of Gaussian states 
    \cite{isar1994open}.
In the case of linear Lindblad operators 
and Hamiltonian $\HQ=\PQ^2/2m+V(\XQ)$, the harmonic approximation
$\LLQLa$ to $\LLQ$ at $\alpha$ is obtained simply by replacing the potential $V$ with its local quadratic approximation about $x=\alpha^\x$. (For the harmonic approximation to more general Lindbladians, see the companion work \cite{hernandez2023decoherence2}.)
This is an open-system generalization of Heller's thawed-Gaussian approximation 
\cite{heller1975time}.
We will find a function $\pas(t)$ that exactly solves
\begin{equation}
\label{eq:rhot-evo-simple-main}
\partial_t \rhot(t) =  
\intNTS
\, \pas(t) \LLQLa [\tauQas].
\end{equation}

In Eq.~\eqref{eq:rhot-evo-simple-main} we evolve each Gaussian state $\tauQas$ in the mixture $\rhot(t)$ according to the harmonic approximation to the dynamics centered at $\alpha$.  The form of Eq.~\eqref{eq:simple-trajectories} and \eqref{eq:rhot-evo-simple-main} requires that we can continually express $\rhot(t)$ as a mixture of NTS states.  For arbitrary Lindblad dynamics, we could not write $\rhot$ as such a mixture.  For instance, in the limiting case of a closed system ($D=0$), the Gaussian states could become arbitrarily squeezed over time, and these squeezed states could not be rewritten as mixtures of NTS states. We must therefore take advantage of the environment-induced diffusion $\D$ present in the open system, which diffusively broadens each Gaussian state in the mixture, causing it to become mixed, which can then be decomposed into less squeezed Gaussian states satisfying 
$\sigma \le \sigmaNTS$. 
This essential mechanism is illustrated in Figure \ref{fig:ellipses}(b). 

\proofpart{Duhamel bound}
Having defined our approximating trajectory $\rhot$, we can 
compare it to the exact evolution $\rho(t)$ using
the Duhamel formula:
\begin{align}\label{quantum-duhamel-state-main}
\rhot(t)-\rho(t)	& = \int_0^t \!\dd s \,  e^{ (t-s) \LLQ} \left( \partial_s - \LLQ \right) [\rhot(s)].
\end{align}
This allows us to express the error at time $t$ as an accumulation of errors at previous times, yielding
\begin{align}\label{eq:quantum-bound-special-case-with-max-main}
\Trnorm{\rhot(t)-\rho(t)}
\leq t 
\sup_{\sigma \le \sigmaNTS}
\sup_{\alpha} \Trnorm{ (\LLQLa-\LLQ) [\tauQas]}.
\end{align}
The corresponding classical expression is similar:
\begin{align}\label{eq:classical-bound-special-case-with-max-main}
\|\tilde{\cstate}(t)-\rhoc(t)\|_{L^1} \le t 
\sup_{\sigma \le \sigmaNTS}
\sup_{\alpha}
\Lonenorm{ (\LLCLa-\LLC) [\tauCas]}.
\end{align}
Now we only need to bound the errors appearing above, involving the difference between the exact evolutions $\LLQ$ and $\LLC$ and the corresponding harmonic approximations $\LLQLa$ and  $\LLCLa$. 

\proofpart{Harmonic approximation error}

The error from the harmonic approximation will be small when the spatial size of the Gaussian state is small compared to the anharmonicity of the potential $V$.  Indeed, the only modification necessary to obtain the harmonic approximation to the dynamics involves replacing the potential $V$ with its second-order Taylor approximation at $x=\alpha^\x$, so the error is proportional to the max leading-order (i.e., third-order) correction $\Delta x^3\JkSN{V}{3}$ at the characteristic spatial width $\Delta x\sim\|\sigma^{\x\x}\|_{\mathrm{op}}^{\frac{1}{2}}$ of the wavepacket with covariance matrix $\sigma$. (The $\x\x$ superscript is used to denote the upper left block of $\sigma$.) More precisely, in Methods Section~\ref{sec:harmonic-error} we prove the bounds
\begin{align}\label{eq:main-trace-harm-err-main}
\Trnorm{ (\LLQLa-\LLQ)[\tauQas] }  &\le \er_\sigma,
\quad
\Lonenorm{(\LLQLa-\LLQ) [\tauCas]} \le  \er_\sigma
\end{align}
where 
\begin{align}\label{err-rate}
\er_\sigma = \sqrt{3} \, d^{\frac32} \hbar^{-1}\|\sigma^{\x\x}\|_{\mathrm{op}}^{\frac{3}{2}} \JkSN{V}{3}.
\end{align}
See Figure \ref{fig:harm-err} for a visual derivation of this harmonic error term.  
In the quantum case, the  $\hbar^{-1}$ factor above can be seen to arise from the $\hbar^{-1}$ in the Schr{\"o}dinger equation; in the classical case, it is related to the fact that the coherent state has area $\hbar$ in phase space.  

The harmonic errors appearing in the quantum error \eqref{eq:quantum-bound-special-case-with-max-main} and classical error \eqref{eq:classical-bound-special-case-with-max-main} are taken only over covariance matrices satisfying $\sigma\le \sigmaNTS = \zz\sigmaco$. Because $\|\sigma^{\x\x}\| \le \zz\|\sigmaco^{\x\x}\| = \zz\hbar/2\unitRatio_H$, we can combine 
(\ref{eq:quantum-bound-special-case-with-max-main}--\ref{err-rate})
to get the error rate bound \eqref{eq:bound-simple-potential-eps}.

\section{Physical examples}
\label{sec:physical-examples}

To compare with previous studies, let us estimate the strength of the environment-induced diffusion for real physical systems: a grain of dust being decohered by the cosmic microwave background, a large molecule being decohered by sunlight, and so on \cite{joos2013decoherence}.  Many of the strongest sources of decoherence act much more strongly on position than momentum 
\cite{paz1993reduction},
i.e., the diffusion matrix is approximately degenerate: $\D \approx \diag(0,\Dp)$. This makes our error bound \eqref{eq:bound-simple-potential} weak, per \eqref{eq:bound-simple-potential-eps} and \eqref{eq:D0}.  However, because the error rate \eqref{err-rate} depends only on the position block $\sigma^{\x\x}$ 
of the covariance matrix of the Gaussian states used in the construction of $\rhot$, we can strengthen the theorem by allowing any pure Gaussian state with covariance matrix satisfying $\sigma^{\x\x} \le (\hbar^2/2\Dp \tH )\IdM_d$ or, equivalently, $\sigma^{\p\p} \ge (\Dp \tH /2)\IdM_d$. We can call these states \introduce{not too thin} (NTT), since they are a superset of the NTS state where the squeezing constraint is enforced along only the position directions of phase space.  The strengthened theorem then bounds the quantum-classical error rate \eqref{eq:bound-simple-potential} as 
\begin{align}
\er 
&= d^{\frac32}  \hbar^2 m^{-\frac34} \JkSN{V}{2}^{\frac{3}{4}}  \JkSN{V}{3} (\Dp)^{-\frac32}.
\end{align} 
when the initial state is NTT.
(Although in principle the NTT constraint is compatible with the state $\rhot(t)$ accumulating coherence between arbitrarily different amounts of momentum, it's previously been observed 
\cite{paz1993reduction}
that the $\hat{p}^2/2m$ part of the Hamiltonian smears that momentum coherence into the position direction where it is then decohered.)

In other words, it will not be feasible to distinguish the quantum and classical dynamics  observables by measuring any observable for a time
\begin{align}\label{eq:example-breaking-time}
T \sim \er^{-1} =
\frac{x_H\tH^{\frac72}(\Dp)^{\frac32}}{d^{\frac32} \hbar^2 m}
\end{align} 
where $\tH = \sqrt{m/\JkSN{V}{2}}$ and $x_H = \JkSN{V}{2}/\JkSN{V}{3}$ are the characteristic time and length scales of the dust particle's motion.

For instance, a dust particle with diameter  $10\, \mathrm{\mu m}$ and mass
$m\sim 10^{-11}$ kg decohering 
in sunlight ($T\sim 5800$ K) has a momentum diffusion rate $\Dp = \hbar^2 \Lambdax \sim 10^{-43} \mathrm{kg}^2 \mathrm{m}^{2}\mathrm{s}^{-3}$ 
(Table 3.1 of Ref.~\cite{joos2013decoherence}).
If it moves through a macroscopic potential varying on the length scale of $x_H \sim 1\, \mathrm{m}$ at characteristic speed $x_H/\tH \sim 1 \mathrm{m/s}$, the errors $\Trnorm{\rho(t)-\rhot(t)}$ and $\|\tilde{f}(t)-\rhoc(t)\|_{L^1}$ are small for 
$T \sim  2$ million years, in comparison to the Ehrenfest time $\teh \sim \lambdaL^{-1}\log(s_H/\hbar) \sim 1$ minute.  
The timescale needed for the sunlight to appreciably change the energy of the dust particle, which would otherwise be conserved, is 
$(m x_H/\tH)^2/\Dp \sim 30$ trillion years, 
so its classical motion appears noiseless and reversible.

\vspace{1.5em}
\section{Acknowledgements}
We thank Yonah Borns-Weil for valuable discussion and collaboration on related efforts, Eugene Tang for discussion of the manuscript, Edwin Ng for numerical simulation techniques, and Wojciech Zurek for foundational insights that inspired this work. We thank Joseph Emerson, Fabricio Toscano, and Diego Wisniacki for helpful discussion of numerical evidence for the quantum-classical correspondence in closed and open systems.  DR acknowledges funding from NTT Research (Grant AGMT DTD 9/24/20).

\bibliographystyle{unsrt}
\bibliography{references}

\end{multicols}

\newpage
\appendix
\begin{center}{\LARGE \textbf{Methods}}\end{center}
\vspace{1em}
\vspace{2em}

Our methods are organized as follows.  In Section~\ref{app:norm-interp}, \ref{sec:wigner}, and \ref{sec:symp-ham} we recall basic definitions and properties of some norms, the Wigner function, and symplectic and Hamiltonian matrices.  In Section~\ref{sec:proof} we prove Theorem \ref{thm:mainResult}, making use of Lemma~\ref{lem:HarmErrorQV} bounding the harmonic approximation error, which in turn is proved in Section~\ref{sec:harmonic-error}. 
In Section~\ref{sec:simulation} we prove Theorem~\ref{thm:simulation} regarding the computational costs of simulating open quantum systems.

\section{Norms}\label{app:norm-interp}

For a function $f(\alpha)$ over phase space variable $\alpha = (x,p) \in \mathbb{R}^{2d}$, the \introduce{(Lebesgue) $L^q$ norm} is
\begin{align}
\norm{f}_{L^q} := \left( \int\! \dd \alpha \left|f(\alpha)\right|^q\right)^{1/q}.
\end{align}
Our classical error bound on the difference in two distributions $f$ and $g$ is stated with $L^1$ norm: $\norm{f-g}_{L^1} := \int \! \dd \alpha |f(\alpha)-g(\alpha)|$. The analogous norm on the quantum side is the \introduce{trace norm} $\|\hat{A}\|_{\mathrm{Tr}} = \Tr[(\hat{A}^\dagger \hat{A})^{1/2}] = \sum_i \sigma_i(\hat{A})$ of an operator $\hat{A}$, i.e., the sum of the singular values $\sigma_i(\hat{A})$.  

The importance of these two norms follows from their standard variational characterization. For quantum states $\rho$ and $\rhot$, the respective probabilities $q_\qu=\Tr[\hat{Q}\rho(t)]$ and $\tilde{q}_\qu=\Tr[\hat{Q}\rhot]$ for the measurement outcome associated with a projector $\hat{Q}\le \IdQ$ are constrained by 
\begin{align}\begin{split}\label{eq:q-norm-error}
|q_\qu-\tilde{q}_\qu| &= |\Tr[Q(\rho - \rhot)]| \\
&\le \|\hat{Q}\|_{\mathrm{op}} \Trnorm{\rho - \rhot} \\
& \le \Trnorm{\rho - \rhot}
\end{split}\end{align}
Likewise, for classical states $\cstate$ and $\cstatet$ the respective probabilities $q_\cl=\int\!\dd\alpha\,Q(\alpha)\cstate(\alpha)$ and $\tilde{q}_\cl=\int\!\dd\alpha\,Q(\alpha)\cstatet(\alpha)$ for the measurement of any classical indicator variable $Q(\alpha)\le 1$ are constrained by
\begin{align}\begin{split}\label{eq:c-norm-error}
|q_\cl-\tilde{q}_\cl| &= \left|\int\!\dd\alpha\,Q(\alpha)(\cstate(\alpha)-\cstatet(\alpha))\right| \\
&\le \|Q\|_{L^\infty} \|\cstate - \cstatet\|_{L^1} \\
& \le \|\cstate - \cstatet\|_{L^1}
\end{split}\end{align}
Thus, two classical states cannot be readily distinguished when they are close in $L^1$ norm, and two quantum states cannot be readily distinguished when they are close in trace norm, no matter what measurement is performed.

\section{Wigner function}
\label{sec:wigner}

The Weyl representation and the closely related Wigner function are not necessary for stating our main result, Theorem~\ref{thm:mainResult}, but in its proof and in Corollary~\ref{cor:Main} we make use of these three facts:
\begin{enumerate}
\item When $\rho$ is a Gaussian state, its Wigner function $\WW[\rho]$ is a Gaussian probability distribution on phase space with the same mean and covariance: $\WW[\tauQas] = \tauCas$.
\item When $\rho$ is a mixture of Gaussian states, $\WW[\rho]$ is a mixture of the respective Gaussian distributions, because $\WW[\cdot]$ is a linear map.
\item The Weyl trace formula in the special case of the expectation value of a function $V$ of position for a Gaussian state is
\begin{align}
\Tr[\tauQas V(\XQ)] = \int \! \dd\alpha \, \tauCas(\alpha) V(\alpha^\x).
\end{align}
\end{enumerate}
The companion paper, Ref.~\cite{hernandez2023decoherence2}, more thoroughly discusses the Weyl representation in the context of these and related results. For more about the Wigner function in general, see e.g.\ the review in \cite{curtright2014concise}.

\section{Symplectic and Hamiltonian matrices} \label{sec:symp-ham}
In this section we recall basic facts about symplectic and Hamiltonian matrices.  Our starting point is the 
symplectic form
\begin{align}
\sf = \begin{pmatrix}
0 & \IdM_d \\
-\IdM_d & 0 
\end{pmatrix}
\end{align}
associated with the phase space $\bbR^{2d}$ for $d$ classical degrees of freedom.
A \introduce{symplectic} matrix $A$ satisfies
\begin{equation}
A^\tp \sf A = \sf.
\end{equation}
Note that the symplectic matrices $\mathrm{Sp}(2d,\mathbb{R})$ form a group which is a subgroup of the special linear
group $\mathrm{SL}(2d,\mathbb{R})$.

For a one-parameter family $A=A(t)$ of symplectic matrices, 
\begin{equation}
\label{eq:time-derivative}
\begin{split}
0 &= \frac{\dd}{\dd t} (A^\tp \sf A) \\
&= \dot{A}^\tp \sf A + A^\tp \sf \dot{A}.
\end{split}
\end{equation}
Taking $A(t)=\exp(tF)$ so $\dot{A} = FA = AF$, we see that the symplectic group $\mathrm{Sp}(2d,\bbR)$ is generated by the Lie algebra $\mathrm{sp}(2d,\bbR)$, i.e., the set of Hamiltonian matrices $F$ defined by the constraint
\begin{equation}
F^\tp \sf + \sf F = 0.
\end{equation}
[Hamiltonian \emph{matrices} should not to be confused with the classical Hamiltonian variable (the energy) or the corresponding quantum Hamiltonian operator.]
Using $\sf^2 = \IdM_{2d}$ and $\sf^\tp = -\sf$, we can rearrange this as
\begin{equation}
F^\tp = \omega F \omega = -\sf^\tp F \sf.
\end{equation}
Using the identities $\sf = A^\tp \sf A = A^{-\tp}\sf A^{-1}$
we can see that if $F$ is Hamiltonian and $A$ is symplectic, then 
$A^{-1}FA$ is also symplectic.

We can also rearrange~\eqref{eq:time-derivative}
by left-multiplying by $A^{-\tp}$ and right-multiplying
by $A^{-1}$:
\begin{equation}
A^{-\tp} \dot{A}^\tp \sf + \sf \dot{A} A^{-1} = 0.
\end{equation}
Thus $\dot{A}A^{-1}$ is Hamiltonian when $A=A(t)$ is symplectic.
Conjugating by $A$ we can see that $A^{-1}\dot{A}$ is also Hamiltonian
in this case. 

When $A$ is additionally symmetric, 
it follows that $\dot{A}$ is also symmetric so we can write 
\begin{equation}
\dot{A} = FA + A  F^\tp  
\end{equation}
for some Hamiltonian matrix $F$.  

In order that $\sigma$ is the covariance matrices of a pure Gaussian state, it must be positive definite ($\sigma > 0$), and $\sigma/(\hbar/2)$ must be symplectic ($\sigma^\tp \sf \sigma = (\hbar/2)^2\sf$) \cite{simon1994quantumnoise,weedbrook2012gaussian}.  As can be checked, this is true for 
\begin{align}
\sigmaco := \frac{\hbar}{2}\left(\begin{matrix} \unitRatio_H^{-1} \IdM_d & 0 \\ 0 & \unitRatio_H \IdM_d \end{matrix}\right).
\end{align}
In the proof of Theorem~\ref{thm:mainResult} we work with such covariance matrices $\sigma$ that additionally satisfy
the matrix inequality
\begin{align}
\label{eq:sigma-ub}
\sigma \leq \zz\sigmaco
\end{align}
for a particular $\zz\ge 1$.  Inverting this (and using the positivity of $\sigma$ and $\sigmaco$) we obtain the inequality 
\begin{equation}
\label{eq:sigmainv-lb}
\sigma^{-1} \geq \zz^{-1}\sigmaco^{-1}.
\end{equation} 
Now rewriting the symplectic conditions on $\sigma/(\hbar/2)$ and $\sigmaco/(\hbar/2)$ as
\begin{align}
\label{eq:alt-symplectic-1}
\sigma^{-1} &\,= (\hbar/2)^{-2}\sf^\tp\sigma\sf,\\
\label{eq:alt-symplectic-2}
\sigmaco^{-1} &\,= (\hbar/2)^{-2}\sf^\tp\sigmaco\sf,
\end{align}
we derive from~\eqref{eq:sigmainv-lb} the inequality
\begin{equation}
\sf^\tp\sigma\sf \ge \zz^{-1}\sf^\tp\sigmaco\sf.
\end{equation}
De-conjugating by the symplectic form $\sf$ gives our desired equivalent NTS condition:
$\sigma \ge \zz^{-1}\sigmaco$.  That is, we conclude 
\begin{equation}\label{eq:NTL-NTT}
\sigma \ge \zz^{-1}\sigmaco \iff     \sigma \leq \zz\sigmaco
\end{equation}
for covariance matrices of pure Gaussian states. 
More generally, for a pure Gaussian state, the eigenvalues of the covariance matrix $\sigma$ come in pairs $\lambda$ and $(\hbar/2)^2\lambda^{-1}$ \cite{simon1994quantumnoise,weedbrook2012gaussian}.

\section{Proof of Theorem \ref{thm:mainResult}}\label{sec:proof}

Here we prove our main result.  (Our companion work~\cite{hernandez2023decoherence2} addresses a broader class of dynamics and includes technical discussion about the trace- and positivity-preservation of $\LLQ$. See Ref.~\cite{li2024long} for  trace-norm bounds directly between $\rho(t)$ and $\WW^{-1}[f(t)]$, and see Ref.~\cite{galkowski2024classical} for bounds with the Hilbert-Schmidt norm.)

\begin{proof}[Proof of Theorem \ref{thm:mainResult}]
We will build a trajectory $\rhot(t)$ that approximates the true evolution $\rho(t)$ by using a mixture of Gaussian states $\tauQas$, each centered at a point  $\alpha = (\alpha^\x,\alpha^\p)\in\bbR^{2d}$ in phase
space with covariance matrix $\sigma \in\bbR^{2d} \times \bbR^{2d}$,
\begin{align} \label{eq:rhot_pfunc}
\rhot(t)= \iint\! \dd \alpha  \dd \sigma  \, \pas(t) \tauQas,
\end{align}
where $\pas(t)$ only supports covariance matrices $\sigma$ of pure states that are ``not too squeezed'' (NTS) in the sense 
$\sigma \le \sigmaNTS$. 
We assume the true initial state $\rho(t\liq 0) = \iint\! \dd \alpha \dd \sigma  \, \pas(t\liq 0) \tauQas,$ is a (possibly trivial) mixture of such states, so that the true trajectory and our approximation initially coincide: $\rhot(t\liq 0)=\rho(t\liq 0)$. 

We divide the proof into three steps.

\proofpart{Constructing the distribution $\pas(t)$}
We use  $\LLQLa$ to denote the harmonic approximation to $\LLQ$ with respect to the point $\alpha$ in phase space. Then $\LLQLa$ is obtained from $\LLQ$ simply by replacing the potential with its local quadratic approximation about $x=\alpha^\x$, i.e., by 
replacing $V(\XQ)$ with $V^{[\alpha^\x,2]}(\XQ)$ where
$V^{[\alpha^\x,2]} (x) := V(\alpha^\x) + (x-\alpha^\x)\cdot\nabla V(\alpha^\x) + [(x-\alpha^\x) \cdot \nabla]^2 V(\alpha^\x)$  is the second-order Taylor approximation about $\alpha^\x$.
Harmonic dynamics (i.e., quadratic Hamiltonian and linear Lindblad operators) exactly preserve the set of Gaussian states \cite{isar1994open,brodier2004symplectic,graefe2018lindblad}. 
We will find a function $\pas(t)$ that \emph{exactly} solves
\begin{equation}
\label{eq:rhot-evo-simple}
\partial_t \rhot(t) =  \iint\! \dd \alpha \dd \sigma  \, \pas(t) \LLQLa [\tauQas].
\end{equation}
Before we proceed to analyze the error between
$\rhot(t)$ and $\rho(t)$, we first show that 
a function $\pas(t)$ solving~\eqref{eq:rhot-evo-simple} can be found.

Under the harmonic approximation about $\alpha$, a Gaussian evolves such that \cite{isar1994open,brodier2004symplectic,graefe2018lindblad}  its centroid $\alpha$ follows the classical (diffusionless) flow on phase space with flow vector 
\begin{align}
\label{eq:special-case-adot}
\adot(\alpha) :=&  
\left(\begin{matrix}  \alpha^\p/m \\ -\nabla V(\alpha^\x)  \end{matrix}\right)
\end{align}
in the sense that $\dd\alpha(t)/\dd t = \adot(\alpha)$, while the covariance matrix $\sigma$ evolves by $\dd\sigma(t)/\dd t = \sdot(\alpha,\sigma)$ where 
\begin{align}
\label{eq:special-case-sdot}
\sdot(\alpha,\sigma) &:= \hh(\alpha)\sigma+\sigma\hh^\tp(\alpha) + \D 
,
\\
\label{eq:special-case-f}
\hh(\alpha) &:=   
\left(\begin{matrix}  0 & -\IdM_d /m \\ \nabla^2 V (\alpha^\x)  &  0  \end{matrix}\right),
\end{align}
and where $\D = \diag(\Dx\IdM_d,\Dp\IdM_d)$ is the diffusion matrix. This describes the skewing (by $\hh$) and broadening (by $\D$) of the Gaussian; only the latter increases mixedness.  Equivalently, the Gaussian state obeys 
\begin{align}\label{eq:harmonic-gaussian-evo}
\LLQLa [\tauQas]
&= \left[\adot(\alpha) \cdot \partial_\alpha  + \sdot(\alpha,\sigma) \cdot \partial_\sigma \right] \tauQas.
\end{align}

With \eqref{eq:rhot_pfunc} and \eqref{eq:harmonic-gaussian-evo}, our desired condition~\eqref{eq:rhot-evo-simple} becomes 
\begin{align}
\label{eq:pas-weak-eq}
\begin{split}
	& \hspace{-1em} \iint \! \dd \alpha \dd\sigma \, \tauQas \frac{\dd}{\dd t}\pas(t) 
	\\
	& 
	= \iint \! \dd\alpha \dd\sigma \, \pas(t) 
	\left[\adot(\alpha) \cdot \partial_\alpha  + \sdot(\alpha,\sigma) \cdot \partial_\sigma \right]
	\tauQas.
\end{split}
\end{align}

We could integrate the right-hand side of~\eqref{eq:pas-weak-eq} by parts in $\sigma$ and $\alpha$ to obtain a transport equation for \emph{one} solution $p_{\alpha,\sigma}$, but we would quickly lose control of the covariance matrix $\sigma$, which could be stretched arbitrarily long by the evolution. Instead, we observe that any component of the flow in the ``positive'' $\sigma$ direction (which increases mixedness of the state) can also be re-interpreted as diffusion in the $\alpha$ direction \cite{weidlich1967quantumI, weidlich1967quantumII, haken1967quantum, lax1967quantum, isar1991quasiprobability, diosi2000robustness}.

For any choice of decomposition $\sdot=\sdotD+\sdotZ$ 
we have
$\sdot \cdot\partial_\sigma
\tauQas = 
[\sdotZ \cdot\partial_\sigma 
+ \frac{1}{2} \sdotD \cdot\partial_\alpha\partial_\alpha ]
\tauQas$
by the Gaussian derivative identity $\partial_{\sigma} \tauQas = \frac{1}{2} \partial_{\alpha}  \partial_{\alpha} \tauQas$ (reviewed in 
the Supplementary Information).
Plugging this into~\eqref{eq:pas-weak-eq} and integrating by parts,
we see that~\eqref{eq:rhot-evo-simple} is satisfied (with $\pas$ guaranteed to be non-negative when $\sdotD \geq 0$) so long as 
$\pas(t)$ solves $\partial_t \pas(t) = \LLPa [\pas(t)]$ with
\begin{align}
	\label{eq:pas-evolution-eq}
	\begin{split}
		\LLPa[\pas] :=&\, \Big[-\partial_\alpha  \cdot\UC(\alpha) -\partial_\sigma \cdot \sdotZ(\alpha,\sigma)
		+ \frac{1}{2} \partial_\alpha\partial_\alpha \cdot\sdotD(\alpha,\sigma)\Big]\pas
	\end{split}
\end{align}
where the partial derivatives in \eqref{eq:pas-evolution-eq} are understood to act also on $\pas$. (We will show below that the support of $p_{\alpha,\sigma}$ is contained within the set of NTS states, so that there is no boundary term when integrating by parts.)

We have some limited freedom in choosing $\sdotZ$ and $\sdotD$; that is, the decomposition of $S$ into skewing and broadening parts is not unique.  We require (i) $\sdotD\geq0$ so that $\pas$ undergoes non-negative diffusion, and we must choose $\sdotZ$ so that $\pas$ remains supported on covariance matrices of (ii) pure Gaussian states that (iii) satisfy the NTS condition $\sigma \le z \sigmaco$.  Below, we choose a decomposition fulfilling these three requirements.

It is useful to work with the rescaled matrices
$\tild{X}:=\sigmaco^{-1/2}X\sigmaco^{-1/2}$ for $X = \sigma, \sdot, \sdotZ, \sdotD, \D$.  We also use the asymmetric $\tildalt{\hh}:=\sigmaco^{-1/2}\hh\sigmaco^{1/2}$ 
(note $\pm \, 1/2$ exponents) 
so that $\tildsdot = \tildalt{\hh}\tild{\sigma} + \tild{\sigma}\tildalt{\hh}^\tp+\tild{\D}$. 
Then we make the choice
\begin{align}
	\label{eq:sdotZ-choice}
	\tildsdotZ(\alpha,{\sigma}) &:= 
	[\tildalt{\hh}(\alpha) - \platonicm(\tild{\sigma})]\tild{\sigma} + 
	\tild{\sigma}[\tildalt{\hh}^\tp(\alpha) - \platonicm(\tild{\sigma})],
	\\
	\label{eq:sdotD-choice}
	\tildsdotD(\alpha,{\sigma}) &:= \tild{\D}
	+ [\platonicm(\tild{\sigma})\tild{\sigma} + \tild{\sigma} \platonicm(\tild{\sigma})],
\end{align}
satisfying $\tildsdotZ + \tildsdotD = \tildsdot$, 
where
\begin{equation}
	\begin{split}
		\platonicm(\tild{\sigma}) &=
		\left(\frac{\Dz s_H}{\hbar \tH}\right)\frac{\tild{\sigma} - \tild{\sigma}^{-1}}{1-\zz^{-2}}.
	\end{split}
\end{equation}

First, a bit of algebra shows that $\tildsdotZ = \tildsdotZ{}^\tp$ and $(\tild{\sigma}^{-1}\tildsdotZ)^\tp = -\sf^\tp\tild{\sigma}^{-1}\tildsdotZ\sf$ (because $\sf^\tp \tild{\sigma}\sf = \tild{\sigma}^{-1}$). As recalled in Appendix \ref{sec:symp-ham}, this ensures that a covariance matrix evolving by $\sdotZ$ remains a covariance matrix for a pure Gaussian state under the dynamics \eqref{eq:pas-evolution-eq}. 

Next, we need to show that with these choices the distribution $\pas(t)$ never develops support on covariance matrices violating the NTS condition $\sigma \le \zz\sigmaco$, i.e., that $\tild\sigma = \sigmaco^{-1/2}\sigma\sigmaco^{-1/2} \le \zz\IdM_d$ for any $\sigma$ such $\pas \neq 0$.  Note that an equivalent NTS condition is $\sigma \ge \zz^{-1} \sigma_*$ because $\sigma/(\hbar/2)$ and $\sigmaco/(\hbar/2)$ are symplectic matrices. (See Appendix \ref{sec:symp-ham}, Eq.~\eqref{eq:NTL-NTT} for an elementary demonstration.) 
Recall that when symmetric matrix $\sigma$ has distinct eigenvalue $\lambda$ with eigenvector $v$, then $\dot{\lambda} = \langle v | \dot{\sigma} |v\rangle$.
Then the lower-bound condition above will be preserved if $\langle v| \tildsdotZ(\alpha,\sigma) | v\rangle \geq 0$ whenever $v$ is an eigenvector of $\tild{\sigma}$ with eigenvalue $\lambda \leq \zz^{-1}$ ($\tild{\sigma} v = \lambda v$).   
From \eqref{eq:sdotD-choice} we compute
\begin{equation}
	\begin{split}
		\langle v |\tildsdotZ(\alpha,\sigma)|v\rangle 
		&\geq -2\|\tildalt{\hh}(\alpha)\|_{\mathrm{op}}\zz^{-1} +
		2 \Dz s_H \tH^{-1} \ge 0
	\end{split}
\end{equation}
because $\zz :=\max\{\hbar/s_H \Dz,1\} \ge \hbar/s_H \Dz$ and $\|\tildalt{\hh}(\alpha)\|_{\mathrm{op}} \le \sqrt{\JkSN{V}{2}/m} = \tH^{-1}$ by \eqref{eq:special-case-f}. (Here, $\|\,\cdot\,\|_{\mathrm{op}}$ denotes the operator norm, i.e., the largest singular value of a matrix.) 

This ensures that $\pas(t)$ is only supported on NTS covariance matrices ($\zz^{-1} \sigmaco \le \sigma \le \zz\sigmaco$) for all time. 
Then using $\tild{\D} =\sigmaco^{-1/2}\D\sigmaco^{-1/2} = 2s_H \D_H^{-1/2}\D\D_H^{-1/2}/(\tH\hbar) \ge 2\Dz s_H\IdM_{2d}/(\hbar\tH)$ we have by \eqref{eq:sdotD-choice} that $\tildsdotD(\alpha,\sigma) \ge 0$ for all allowed $\sigma$ so $\sdotD(\alpha,\sigma) = \sigmaco^{1/2}\tildsdotD(\alpha,\sigma)\sigmaco^{1/2} \ge 0$. 
Therefore the diffusion term in \eqref{eq:pas-evolution-eq} ensures that $\pas(t)\ge 0$ for all $t\ge 0$, i.e., $\rhot(t)$ is always a true mixture of squeezed Gaussians.

To summarize, we have constructed the trajectory $\rhot(t)$ defined through \eqref{eq:rhot-evo-simple} with the probability distribution $\pas(t)$ supported on pure-state covariance matrices satisfying the NTS condition 
$\sigma\le \zz \sigmaco$ 
defined through the dynamics $\partial_t \pas(t) = \LLPa [\pas(t)]$ of \eqref{eq:pas-evolution-eq} using choices \eqref{eq:sdotZ-choice} and \eqref{eq:sdotD-choice}.

\proofpart{Duhamel bound}
Having defined our trajectory $\rhot$, we can 
compare it to the exact evolution $\rho(t)$ using
the Duhamel formula:
\begin{align}\label{quantum-duhamel-state}
	\rhot(t)-\rho(t)	& = \int_0^t \!\dd s \,  e^{ (t-s) \LLQ} \left( \partial_s - \LLQ \right) [\rhot(s)]
\end{align}
Then
\begin{align}
	& \hspace{-1em} \Trnorm{\rhot(t)-\rho(t)} \\
	& = \Trnorm{  \int_0^t \!\dd s  e^{ (t-s) \LLQ} \left( \partial_s - \LLQ \right) [\rhot(s)] } \\
	& \le  \int_0^t \!\dd s   \Trnorm{ e^{ (t-s) \LLQ} \left( \partial_s - \LLQ \right) [\rhot(s)] } \\
	\label{eq:dog0}
	& \le  \int_0^t \!\dd s   \Trnorm{ \left( \partial_s - \LLQ \right) [\rhot(s)] } \\
	\label{eq:dog1}
	& =  \int_0^t \!\dd s   \Trnorm{ \iint\! \dd \alpha  \dd \sigma \, \pas(s) \left(  \LLQLa - \LLQ \right)    [\tauQas] } \\
	\label{eq:dog2}
	&\le \int_0^t \!\dd s \iint\! \dd \alpha  \dd \sigma\,  \pas(s)   \Trnorm{ \LLQRa [\tauQas] }
	\\
	\begin{split}
		&\le \sup_{\sigma \le \zz\sigmaco}\sup_{\alpha}  \Trnorm{ \LLQRa [\tauQas]}  \int_0^t \! \dd s 
	\end{split}
	\\
	\label{eq:quantum-bound-special-case-with-max}
	&= t \sup_{\sigma \le \zz\sigmaco}\sup_{\alpha} \Trnorm{ \LLQRa [\tauQas]},
\end{align}
where  \eqref{eq:dog0} follows from the fact that $e^{ (t-s) \LLQ}$ is a completely positive map and so cannot increase the trace norm, \eqref{eq:dog1} follows from \eqref{eq:rhot-evo-simple} and the normalization of $\pa(t)$, and in \eqref{eq:dog2} we have defined $\LLQRa:=\LLQ - \LLQLa$.

If we make the replacements $\rhot\to\WW[\rhot] = \tilde{\cstate}$, $\rho\to\rhoc$, $\tauQas\to\tauCas$, $\LLQ\to\LLC$, and $\Trnorm{\,\cdot\,}\to \Lonenorm{\,\cdot\,}$ in (\ref{quantum-duhamel-state}--\ref{eq:quantum-bound-special-case-with-max}), then identical manipulations give
\begin{align}
	\label{eq:classical-bound-special-case-with-max}
	\Lonenorm{\tilde{\cstate}-\rhoc(t)} \le t 
	\sup_{\sigma \le \zz\sigmaco}\sup_{\alpha}
	\Lonenorm{ \LLCRa [\tauCas]}.
\end{align}
Here $\LLCRa:=\LLC - \LLCLa$ where, analogously to $\LLQLa$, $\LLCLa$ is the harmonic approximation to the classical Liovillian obtained by replacing $V\to V^{[\alpha^\x,2]}$ in \eqref{eq:fp-simple}.  (Evolving a phase-space distribution \emph{pointwise} with $\LLCLa$ results in the so-called truncated Wigner approximation.)

\proofpart{Harmonic approximation error}
In Appendix \ref{sec:harmonic-error}, Lemma~\ref{lem:HarmErrorQV}, we prove the bounds 
\begin{align}\label{eq:main-trace-harm-err}
	\Trnorm{ \LLQRa[\tauQas] }  &\le \er_\sigma,
	\quad
	\Lonenorm{\LLCRa [\tauCas]} \le  \er_\sigma
\end{align}
where 
\begin{align} \label{eq:harm-error-rsigma}
	\er_\sigma := \sqrt{3} \, d^{\frac32} \hbar^{-1}\|\sigma^{\x\x}\|_{\mathrm{op}}^{\frac{3}{2}} \JkSN{V}{3}.
\end{align}

Because $\rhot(t)$ is a mixture of Gaussian states that are NTS, the harmonic errors appearing in the quantum error \eqref{eq:quantum-bound-special-case-with-max} and classical error \eqref{eq:classical-bound-special-case-with-max} are taken only over covariance matrices satisfying $\sigma\le \zz\sigmaco$, so $\|\sigma^{\x\x}\| \le \zz\|\sigmaco^{\x\x}\| = \zz\hbar/2\unitRatio_H$.  
Thus applying \eqref{eq:main-trace-harm-err} we get that quantum and classical error rates are both upper bounded by 
\begin{align}
	\er \le  \sqrt{3}\, d^{\frac32} \hbar^{-1}(\zz\hbar/2\unitRatio_H)^{\frac{3}{2}} \JkSN{V}{3}.
\end{align}
Referring to \eqref{eq:NTSsigma}, we obtain \eqref{eq:bound-simple-potential-eps}.
\end{proof}

A variant of Theorem~\ref{thm:mainResult} with general Hamiltonian and Lindblad operators is addressed in Ref.~\cite{hernandez2023decoherence2}, though the treatment of errors there is both more difficult and less helpfully explicit.  Here we can address a case of intermediate generality, ``separable'' Hamiltonians of the form $\HQ(\XQ,\PQ) = \VP(\PQ) + V(\XQ)$. 
\begin{restatable}[Main result for separable Hamiltonians]
{thm}{thmGeneralVP} 
\label{thm:mainResultSeparable}
Theorem~\ref{thm:mainResult} holds more generally for Hamiltonians of the form $\HQ(\XQ,\PQ) = \VP(\PQ) + V(\XQ)$ where the error rate, Eq.~\eqref{eq:bound-simple-potential-eps}, is doubled to
$\er = 2 d^{\frac32}\tH^{-1}(\hbar/s_H)^{\frac12}\max\left\{(\hbar/s_H)\Dz^{-1}, 1\right\}^{\frac32}$,
using the generalizations $\unitRatio_H :=  \sqrt{\JkSN{V}{2}/\JkSN{\VP}{2}}$, $\tH := 1/\sqrt{\JkSN{V}{2}\JkSN{\VP}{2}}$, 
$s_H^{-\frac12} :=  \tH(\unitRatio_H^{-\frac32} \JkSN{V}{3}+\unitRatio_H^{\frac32} \JkSN{\VP}{3})$, 
$\D_H :=(s_H/\tH) \diag\big( \unitRatio^{-1}_H\IdM_d , \unitRatio_H\IdM_d \big)$, 
and $\Dz := \lambdamin\big(\D_H^{-\frac12}\D \D_H^{-\frac12}\big)$.
\end{restatable}
\noindent The proof is nearly the same as for Theorem~\ref{thm:mainResult}. Eq.~\eqref{eq:will-split-h-for-mod} of
Lemma \ref{lem:HarmErrorQV} (below) is adjusted as
\begin{align}
\Trnorm{\LLQRa[	\tauQas]} 
    &\le \frac{2}{\hbar}\Trnorm{\HQRat \tauQas}
    \le \frac{2}{\hbar}\left(\Trnorm{\delta V^{[\alpha^\x,2]}(\XQ) \tauQas}+\Trnorm{\delta\VP^{[\alpha^\p,2]}(\PQ) \tauQas}\right)
\end{align}
and likewise for $\Lonenorm{\LLCRa[	\tauCas]}$. 
Eq.~\eqref{eq:harm-error-rsigma} then becomes 
\begin{align}
    \er_\sigma = \sqrt{3} \, d^{\frac{3}{2}} \hbar^{-1}\left(\|\sigma^{\x\x}\|_{\mathrm{op}}^{\frac{3}{2}} \JkSN{V}{3} + \|\sigma^{\p\p}\|_{\mathrm{op}}^{\frac{3}{2}} \JkSN{\VP}{3} \right).
\end{align}

\section{Harmonic approximation error bound}\label{sec:harmonic-error}

\begin{restatable}[Harmonic approximation error]{lem}{lemHarmErrorQV}
\label{lem:HarmErrorQV}
For a Lindblad equation \eqref{eq:lindblad-simple} with Hamiltonian $\HQ = \PQ^2/2m+V(\XQ)$ and linear lindblad operators $\LQk$, the error $\LLQRa:= \LLQ - \LLQLa$ for the local harmonic approximation $\LLQLa$ to the quantum dynamics at $\alpha$ acting on the pure Gaussian quantum state $\tauQas$ satisfies
\begin{align}
	\label{eq:quantum-trace-err-further-analysis}
	\Trnorm{ \LLQRa[\tauQas] }  
	&\leq  \sqrt{\frac{5d^3}{3}}    \JkSN{V}{3} \frac{\|\sigma^{\x\x}\|_{\mathrm{op}}^{\frac{3}{2}}}{\hbar}
\end{align}
Likewise for a Fokker-Planck equation \eqref{eq:fp-simple}, the error $\LLCRa:= \LLC - \LLCLa$ for the local harmonic approximation $\LLCLa$ to the classical dynamics at $\alpha$ acting on the Gaussian classical state $\tauCas$ satisfies
\begin{align}\label{eq:classical-l1-err-further-analysis}
	\Lonenorm{\LLCRa [\tauCas]} 
	&\leq  \sqrt{3d^3} \JkSN{V}{3}\frac{\|\sigma^{\x\x}\|_{\mathrm{op}}^{\frac23}}{\hbar} 
\end{align}
\end{restatable}
\begin{proof}
We first bound the quantum error
\begin{align}
	\Trnorm{\LLQRa[	\tauQas]} 
	&= \Trnorm{(\LLQ-\LLQLa)[\tauQas]} \\
	&= \Trnorm{-\frac{i}{\hbar}\left[\HQRat, \tauQas\right]} \\
        \label{eq:will-split-h-for-mod}
	&\le \frac{2}{\hbar}\Trnorm{\HQRat \tauQas}
\end{align}
for the Gaussian quantum state $\tauQas = |\alpha,\sigma\rangle\langle \alpha,\sigma|$ with covariance matrix $\sigma$ and mean $\alpha$. Here, $\IdQ$ denotes the identity operator and $\HQRat = \HQ - \HQLat = \delta V^{[\alpha^\x,2]}(\XQ)$ is the operator error from the harmonic approximation. 
By Taylor's theorem we have the classical remainder from the quadratic approximation
\begin{align}\begin{split}
		\label{eq:hamiltonian-taylor-further-analysis}
		\delta V^{[\alpha^\x,2]}(\alpha^\x + \Delta x) &= V(\alpha^\x+ \Delta x) -  V^{[\alpha^\x,2]}(\alpha^\x+\Delta x) \\
		&= \frac{1}{3!}  \left[(\Delta x \cdot \nabla)^3 V\right](\alpha^\x +\xi \Delta x)
\end{split}\end{align}
for some choice of $\xi\in[0,1]$ (depending on $\alpha^\x$ and $\Delta x$).

	Recalling that $\Trnorm{\ket{\psi}\bra{\phi}}^2 = \|\psi\|^2 \|\phi\|^2$, we have
	\begin{align}
		\Trnorm{ \LLQRa[\tauQas] }^2  & \leq \frac{4}{\hbar^2} \norm{ \HQRat|\alpha,\sigma \rangle }^2\\
		& = \frac{4}{\hbar^2}  \Tr\left[\tauQas( \delta V^{[\alpha^\x,2]}(\XQ))^2\right] \\
		\label{eq:apply-tr-formula}
		& = \frac{4}{\hbar^2} \int \dd \beta \tauCas(\alpha+\beta)\left[\delta V^{[\alpha^\x,2]}(\alpha^\x+\beta^\x)\right]^2
	\end{align}
	where 
	$\tauCas(\alpha+\beta) = \WW[\tauQas](\alpha+\beta) = \exp(-\beta^a\sigma^{-1}_{ab}\beta^b/2)/[(2\pi)^d\sqrt{\det\sigma}]$
	(a positive-valued function on phase space)
	is the Wigner function of the pure Gaussian state $\tauQas=|\alpha,\sigma \rangle\langle\alpha,\sigma|$.
	In Eq.~\eqref{eq:apply-tr-formula} we have made use of the Weyl trace formula in the particularly simple case of the expectation value of an operator that is a function of position: $\Tr[\rho V(\XQ)] = \int\!\dd\alpha\,\WW[\rho](\alpha)V(\alpha^\x)$. 
	Next we apply the approximation \eqref{eq:hamiltonian-taylor-further-analysis} from Taylor's theorem:
	\begin{align}
		\Trnorm{ \LLQRa[\tauQas] }^2  
		&\leq \frac{4}{\hbar^2}  \int \dd \beta \tauCas(\alpha+\beta)
		\left[\frac{1}{3!}[(\beta^\x\cdot\nabla)^3 V](\alpha^\x+\xi(\beta^\x) \beta^\x) \right]^2 
		\\
		\label{eq:q-do-cs}
		& \le \frac{\JkSN{V}{3}^2}{9  \hbar^2}  \int \dd \beta \tauCas(\alpha+\beta) |\beta^\x|^6\\
		& = \frac{\JkSN{V}{3}^2}{9  \hbar^2}  \int \dd \beta \tauCas(\alpha+\beta) (\beta^\tp P_\x \beta)^3
		\\
		\begin{split}
			\label{eq:do-q-g-integral}
			& = \frac{\JkSN{V}{3}^2}{9  \hbar^2}  \Big[(\Tr\sigma^{\x\x})^3+6 (\Tr\sigma^{\x\x})\Tr((\sigma^{\x\x})^2) 
			+8\Tr((\sigma^{\x\x})^3)\Big]
		\end{split}
		\\
		\label{eq:op-norm-bound}
		& \le \frac{\JkSN{V}{3}^2}{9  \hbar^2} \|\sigma^{\x\x}\|_{\mathrm{op}}^3 \left(d^3+6d^2+8d\right)
		\\
		\label{eq:d-q-bound}
		& \le \frac{5\JkSN{V}{3}^2}{3  \hbar^2} \|\sigma^{\x\x}\|_{\mathrm{op}}^3 d^3
	\end{align}
	where $P_\x := \left(\begin{smallmatrix}\IdM_d & 0 \\ 0 & 0 \end{smallmatrix}\right)$ projects onto the $\x$ block. 
	In \eqref{eq:q-do-cs} we have used the Cauchy-Schwartz inequality, in \eqref{eq:do-q-g-integral} we have performed the Gaussian integral (as recalled in
    the Supplementary Information) 
    and in \eqref{eq:op-norm-bound} we have used $\Tr[A^n] \le \|A\|_{\mathrm{op}}^n d$ for $d\times d$ positive semidefinite matrix $A$.
	Eq.~\eqref{eq:op-norm-bound} implies Eq.~\eqref{eq:quantum-trace-err-further-analysis} because $d\ge 1$.
	
	Now we turn to the classical harmonic error:
	\begin{align}
		\Lonenorm{\LLCRa[\tauCas]} & = \Lonenorm{(\LLC-\LLCLa)[\tauCas]} \\
		& = \Lonenorm{-\sf^{ab}\partial_a(\tauCas\partial_b\HCRat)} \\
		& = \Lonenorm{(\partial_\p\tauCas)\cdot(\partial_\x \delta V^{[\alpha^\x,2]})}\\
		& = \Lonenorm{-\tauCas (\sigma^{-1}\beta)_\p \cdot(\partial_\x \delta V^{[\alpha^\x,2]})}\\
		& \leq  \|\tauCas^{1/2}\|_{L^2} 
		\, \|\tauCas^{1/2} (\sigma^{-1}\beta)_\p \cdot(\partial_\x \delta V^{[\alpha^\x,2]})\|_{L^2}
	\end{align}
        where the Gaussian derivative $\partial_a\tauCas = -\sigma^{-1}_{ab} \beta^b \tauCas$ is recalled in 
        the Supplementary Information
        and where in the last line we have used the Cauchy-Schwartz inequality. (Here, $\beta$ inside the norm is understood to represent the linear function $f(\alpha+\beta) = \beta$ for fixed $\alpha$.) 
	Then because $\|\tauCas^{1/2}\|_{L^2}^2  = \int\!\dd \beta\,|\tauCas(\alpha+\beta)| = 1$ we have 
	\begin{align}
		\Lonenorm{ \LLCRa [\tauCas]}^2
		& \leq \int\!\dd \beta\,\tauCas(\alpha+\beta) \left| (\sigma^{-1}\beta)_\p\cdot[(\partial_\x \delta V^{[\alpha^\x,2]})(\alpha^\x+\beta^\x)] \right|^2
		\\
		& = \int\!\dd \beta\,\tauCas(\alpha+\beta) \left| (\sigma^{-1}\beta)_\p \cdot \frac{1}{2!} [(\beta^\x\cdot\partial_\x)^2 \partial_\x V](\alpha^\x+\xi(\beta^\x)\beta^\x)\right|^2
		\\
		& =  \int\!\dd \beta\,\tauCas(\alpha+\beta) \left| \frac{1}{2!} [(\beta^\x\cdot\partial_\x)^2 ((\beta\sigma^{-1})_\p\cdot \partial_\x) V](\alpha^\x+\xi(\beta^\x)\beta^\x)\right|^2
		\\
		& \le  \frac{\JkSN{V}{3}^2}{4} \int\!\dd \beta\,\tauCas(\alpha+\beta) \left[ |\beta^\x|^2 |(\beta\sigma^{-1})_\p|\right]^2
		\\
		& \le  \frac{\JkSN{V}{3}^2}{4} \int\!\dd \beta\,\tauCas(\alpha+\beta) (\beta^\tp P_\x \beta)^2 (\beta^\tp \sigma^{-1} P_\p \sigma^{-1} \beta)
		\\
		& \le  \frac{\JkSN{V}{3}^2}{4}\left[ (\Tr\sigma^{\x\x})^2 \Tr(\sigma^{-1}_{\p\p}) +  2 \Tr(\sigma^{-1}_{\p\p}) \Tr[(\sigma^{\x\x})^2] \right]
		\\
		& = \frac{\JkSN{V}{3}^2}{4}d^2(d+2) \|\sigma^{-1}_{\p\p}\|_{\mathrm{op}} \|\sigma^{\x\x}\|_{\mathrm{op}}^2 \label{eq:L1_harm_error_mixed}
	\end{align}
	where we have again used Taylor's theorem, the Cauchy-Schwartz inequality, and the Gaussian integrals reviewed in 
    the Supplementary Information.

(In particular, $\Tr[\sigma \sigma^{-1} P_\p \sigma^{-1}] = \Tr[\sigma^{-1}_{\p\p}]$ and $\Tr[\sigma P_\x \sigma \sigma^{-1} P_\p \sigma^{-1}] = \Tr[ P_\x  P_\p] = 0$.)
When $\sigma$ is the covariance of a pure Gaussian quantum state, its the eigenvalues come in pairs $\lambda$, $\hbar^2/(4\lambda)$ that are associated with symplectically conjugate directions (Appendix \ref{sec:symp-ham}). This means $\|\sigma^{-1}_{\p\p}\|_{\mathrm{op}} = (4/\hbar^2)\|\sigma_{\x\x}\|_{\mathrm{op}}$ so that
\begin{align}
	\Lonenorm{ \LLCRa [\tauCas]}^2
	\le \, & \frac{\JkSN{V}{3}^2}{\hbar^2}d^2(d+2) \|\sigma^{\x\x}\|_{\mathrm{op}}^3
\end{align}
which implies \eqref{eq:classical-l1-err-further-analysis} because $d\ge 1$. 
\end{proof}

For $N$ particles in $n$ spatial dimensions with $k$-wise interactions, the dimension is $d=Nn$ but the $d^{\frac{3}{2}}$ dimensional factor in Lemma \ref{lem:HarmErrorQV} can be replaced with a factor of only $k^{\frac{3}{2}}n^{\frac{3}{2}}$, 
a more favorable scaling.  One therefore expects extensive error for many-body systems.  This growth in error is analogous to that of the orthogonality catastrophe, and therefore a more local notion of error seems to be needed to study the quantum-classical correspondence in many-body systems.

\section{Classically simulating the Lindblad equation}\label{sec:simulation}

In this section we discuss analytical bounds on the computational complexity of simulating the quantum Lindblad evolution using a classical computer.
While we expect this task to be difficult for closed quantum systems in general, we will see that the simulation becomes tractable when the desired Lindblad equation includes sufficient diffusion.

\subsection{Summary}

Our setup is similar to that Theorem~\ref{thm:mainResultSeparable}, the extension of Theorem~\ref{thm:mainResult} to ``separable'' Hamiltonians. Specifically, we consider Hermitian linear Lindblad functions $L_k$, 
a Hamiltonian of the form $\HQ = \VP(\PQ)+V(\XQ)$ where the second and third derivatives of $V$ and $\VP$ are bounded, and an initial state $\rho(t\liq 0)$ that is a mixture of NTS states as in \eqref{eq:simple-initial-quantum-state}. 

To guarantee a well-behaved numerical simulation, we will additionally demand that the \emph{first} derivatives of $V$ and $\VP$ are bounded.  This ensures that the forces experienced by the system do not become arbitrarily large, which would be problematic for both quantum and classical simulation techniques (e.g., necessitating arbitrarily small Runge-Kutta time steps). Note that, strictly speaking, the usual kinetic energy $\VP(\PQ)= \PQ^2/(2m)$ is therefore forbidden, but one could take $\VP(\PQ)=\frac{1}{2m}\PQ^2 e^{-\PQ / p_0}$ for some momentum scale $p_0$ much larger than the relevant momentum scales of the problem, which resembles $\VP(\PQ)=\PQ^2/(2m)$ for $p \ll p_0$. Alternatively, one could modify our results by allowing unbounded first derivatives of $V$ and $\VP$, and thus admitting $\VP(\PQ)=\PQ^2/2m$, if one proved that the bulk of the distribution remained confined to a bounded region of phase space over the simulation time period (e.g., due to energetic constraints).

Under these assumptions, we ask how the computational complexity of the simulation depends on $\hbar$, the dimensionless scalar diffusion strength $\Dz$ (see Eq.~\eqref{eq:D0}), the final time $T$, and the desired error $\error$.
We will focus on the asymptotic behavior with respect to these parameters and will not keep track of other constants.
The main result is that, as long as the diffusion is strong enough, there is an efficient
classical simulation of the quantum dynamics up to a long time. 

\begin{restatable}[Lindblad Simulation]{thm}{thmSimulation}
    \label{thm:simulation}
    Consider the Lindblad equation~\eqref{eq:lindblad-simple}, with $d$ degrees of freedom, Hamiltonian $\HQ = \VP(\PQ)+V(\XQ)$, linear Hermitian Lindblad operators $\LQk = \ell_{k}\cdot(\XQ,\PQ)$, and initial condition $\rho(t\liq 0)$ satisfying the hypotheses of Theorem \ref{thm:mainResultSeparable}. Assume the first three derivatives of $V$ and $\VP$ are bounded.
    Let $T>0$ be a time and $\error$ be a desired error threshold satisfying the strict inequality $\error > rT$ where
    \begin{align} \label{eq:sim-error-rate}
    	r := 2\frac{d^{\frac32}}{\tH} \sqrt{\frac{\hbar}{s_H}} \max\left\{\frac{\hbar/s_H}{\Dz}, 1\right\}^{\frac32}
    \end{align}
	is a characteristic error rate.
	Then there exists a (randomized) classical algorithm to compute the expectation value of any quantum observable $\hat{A}$ satisfying $\|A\|_{L^\infty} + \|\hat{A}\|_{\mathrm{op}}\leq 1$ within error $\error$
    (with probability at least $99\%$) and time-complexity
    \begin{align}\begin{split}
        \label{eq:sim-complexity}
            &\mathcal{O}\Bigg(
		\mathrm{poly}(d)
		\frac{T/\tH}{(\error-rT)^2}
		\Bigg(
		1+
		\te^{\frac12} \left(
		\frac{1}{\Dz^{\frac14}} +
		\frac{1+\Dinf^{\frac12}}{\Dz^{\frac12}}
		\right) +
		\te  \frac{\Dinf^{\frac12}}{\Dz^{\frac12}}+
		\te^{\frac32} \frac{\Dinf+\Dinf^{\frac32}}{\Dz}  +
		\te^{2} \frac{\Dinf^2}{\Dz} 
		\Bigg) \Bigg),  
    \end{split}\end{align}
where $\Dz,\Dinf = \lambda_{\mathrm{min,max}}(\D_H^{-\frac12} D \D_H^{-\frac12})$ and
where $\te := (T/\tH)/(\error-rT)$.
Here $\mathcal{O}(\cdot)$ hides multiplicative factors that depend on $\JkSN{V}{s}$, $\JkSN{\VP}{s}$ for $s=1,2,3$, but that are independent of $\hbar$, $d$, and  $\D$.
\end{restatable}
\noindent See Section \ref{sec:Lindblad-simulation-proof} for the proof.
Note the error rate threshold $r$ is the same error rate appearing in Theorem \ref{thm:mainResultSeparable}. As the desired error $\error$ approaches $rT$, the time complexity diverges. While we frame the task as computing the expectation value of a single observable, in fact the algorithm does not need to be re-run to compute expectation values of multiple observables.  For $N$ distinct observables, the additional time complexity is an additive term of $N \error^{-2}$.  This is because the algorithm produces $\mathcal{O}(\error^{-2})$ samples of stochastic trajectories, which can then be used to compute expectation values for any observable.

The primary significance of Theorem~\ref{thm:simulation} is the polynomial dependence of the computational complexity,
\begin{align}
	\label{eq:poly-dep}
	\mathcal{O}(\textup{poly}(d,\error^{-1},T/\tH,\Dz^{-1},\Dinf)),
\end{align}
as already highlighted in Section~\ref{sec:classical-simulation-claim} by the (strictly simpler) Theorem~\ref{thm:simulation-simplified}. In particular, the time complexity of the quantum simulation is independent of $\hbar$ and improves as $\Dz$ increases (for small $\Dz$).
This is analogous to the situation in noisy random circuit sampling \cite{aharonov2023polynomial} and noisy boson sampling \cite{qi2020regimes}, in which case the noiseless closed system evolution is hard to simulate but the noisy version is tractable.  Related sampling-based ideas using Wigner functions have appeared before, but they are not directly applicable\footnote{Mari and Eisert \cite{mari2012positive} (also see \cite{veitch2012negative,veitch2013efficient}) use a sampling-based approaching for classically simulating quantum evolution, based on the Wigner function.  However, they only prove efficient simulation of quantum systems when the initial state \emph{and the applied unitary evolution} both have positive Wigner function, using a notion of Wigner functions for superoperators (given by the Wigner function of the associated Choi operator).  It would be interesting to ask whether the Lindblad evolution we consider satisfies an approximate version of this property.}.

Furthermore, we emphasize the polynomial dependence on the $d$ degrees of freedom.  In contrast, for a closed quantum system, one expects that a naive classical simulation requires time-complexity
\begin{align}
	\textrm{poly}(\hbar^{-d}).
\end{align}
We are not aware of any such complexity lower bound in this precise setting, but we can reason as follows.  If the dynamics occur within a fixed volume $S^d$ of phase space, the effective Hilbert space is dimension $S^d/\hbar^d$, which can be made precise.  Then heuristically one would expect that the classical simulation algorithm requires resources polynomial in the effective Hilbert space dimension in the worst case. 

The proof of Theorem~\ref{thm:simulation} uses an algorithm (corresponding to Proposition~\ref{prp:simulation-FP}) for sampling from the density
$f(t)$ solving the Fokker-Planck equation~\eqref{eq:fp-simple}, which we believe to be new, or at least not carefully analyzed before.
That algorithm concerns only classical dynamics and its details are somewhat extraneous to our main results, so its description and the proof of Proposition~\ref{prp:simulation-FP} are confined to the Supplementary Information.  
However, it's not necessary to invoke Proposition~\ref{prp:simulation-FP} in order to see that the complexity is a polynomial. The weaker Lindblad simulation result, Theorem~\ref{thm:simulation-simplified}, can be proven with existing Fokker-Planck simulation algorithms, as shown in Section~\ref{sec:Lindblad-simulation-proof-simplified}.

We suspect that Theorem~\ref{thm:simulation} could be generalized to provide rigorous bounds in cases where diffusion is too weak for strictly classical Fokker-Planck simulations to provide sufficiently accurate estimates. Instead, following known practical techniques, quantum simulation would be deployed within (crucially) bounded \cite{steimle1995mixed,schack1995quantum,wiseman2016quantum, johansson2013qutip} and appropriately squeezed \cite{tezak2017lowdimensional,yanagimoto2022onset} regions of phase space, with appropriately increased complexity cost.

\subsection{Proof of Theorem~\ref{thm:simulation} on simulating the Lindblad equation} \label{sec:Lindblad-simulation-proof}	
	
	Per the assumptions of Theorem~\ref{thm:simulation}.   The Hamiltonian $\HQ = \VP(\PQ)+V(\XQ)$ is separable, $\rho(t)$ is a solution to the Lindblad 
	equation~\eqref{eq:lindblad-simple}, and $f(t)$ is a solution to the corresponding Fokker-Planck equation \ref{eq:fp-simple}.  We consider a desired error $\error$ satisfying the strict inequality $\error > \er T$ where
	we use the appropriate minimum error rate 
	$r$
	from Theorem~\ref{thm:mainResultSeparable}. To estimate $\Tr[\hat{A}\rho(t)]$ we apply the Weyl trace formula 
	$\Tr[\hat{A}\rhot(t)] = \int \! A(\alpha)\WW[\rhot](t)(\alpha)\mathrm{d}\alpha$ 
	to Theorem~\ref{thm:mainResultSeparable} to obtain
	\begin{align}\begin{split}
		\label{eq:expectation-bd}
		\left|\Tr[\hat{A}\rho(T)] - \int \! A(\alpha)f(T,\alpha)\mathrm{d}\alpha\right| 
		&\le \left|\Tr[\hat{A}(\rho(T)-\rhot(T))]\right| +\left| \int \! A(\alpha)(\tilde{f}(T,\alpha)-f(T,\alpha))\mathrm{d}\alpha\right|\\
		&\le (\|\hat{A}\|_{\mathrm{op}}+\|A\|_{L^\infty}) r T \\
		&\le  rT,
	\end{split}\end{align}
	(analogous to Corollary~\ref{cor:Main} of Theorem~\ref{thm:mainResult}.)
	The second line uses the standard expectation value bounds \eqref{eq:q-norm-error} and \eqref{eq:c-norm-error} and the third line uses 
	our hypothesis that $\|A\|_{L^\infty} + \|\hat{A}\|_{\mathrm{op}}\leq 1$.
	Therefore the quantum expectation values $\Tr[\hat{A}\rho(T)]$ can be estimated by computing the classical expectation value 
	$\langle A\rangle_{f(T)} := A(\alpha) f(T,\alpha) \mathrm{d}\alpha$
	which is a problem of sampling points $\alpha$ from the probability distribution $f(t)$
	solving the Fokker-Planck equation.  
	
	To achieve the desired time complexity, we make use of the following theorem about simulating the Fokker-Planck equation, which is proven in the Supplementary Information.
	\begin{restatable}[Fokker-Planck simulation]{prp}{prpSimulationFP}
		\label{prp:simulation-FP}
		Consider the Fokker-Planck equation
		\begin{align}
			\label{eq:fp-sim-repeat}
			\partial_t f = v^a \partial_a f + \frac{1}{2}  D^{ab} \partial_a \partial_b f.
		\end{align}
		where $D^{ab}$ is a constant positive definite matrix and where $v^a$, as vector-valued function over phase space, is bounded and has bounded derivatives up to order $p$ for some integer $p\geq 2$. Assume initial condition $f(t=0)$ may be sampled with $\mathcal{O}(1)$ time-complexity.  
		Then there exists a randomized classical algorithm to sample a distribution that is at most a total variation distance $\auxerror$ from $f(T)$ with time-complexity
		\begin{align}\begin{split}
				\label{eq:sim-complexity-fp}
				& \mathcal{O}\Bigg(
				\mathrm{poly}(d)
				T
				\Bigg(
				\left[
				\left(\frac{T}{\auxerror}\right)^2  \frac{\Dmax}{\Dmin} + 
				\left(\frac{T}{\auxerror}\right)\frac{1}{\Dmin^{1/2}} +
				\left(\frac{T}{\auxerror}\right)^3  \frac{\Dmax^2}{\Dmin^2}+
				\left(\frac{T}{\auxerror}\right)^3 \frac{\Dmax^3}{\Dmin^2} +
				\left(\frac{T}{\auxerror}\right)\frac{1}{\Dmin} +
				\left(\frac{T}{\auxerror}\right)\frac{\Dmax}{\Dmin}
				\right]^{1/p} \\
				&\qquad\qquad\qquad\qquad\qquad +
				\left(\frac{T}{\auxerror}\right)^{2} \frac{\Dmax^2}{\Dmin} +
				1
				\Bigg) \Bigg),  
		\end{split}\end{align}
		where  $\mathcal{O}(\cdot)$ hides a multiplicative coefficients depending 
		on $p$ and $\JkSN{v^a}{s}$ for $0\le s \le p$ (and $a = \x,\p$).
	\end{restatable}  
	
	We want to invoke Proposition~\ref{prp:simulation-FP} with $v^a =\partial_a H$ and $p=2$.   But to do so, we need to convert the Fokker-Planck equation from Theorem~\ref{thm:mainResultSeparable} into a dimensionless form so that taking the minimum and maximum eigenvalues of $\D$ in Proposition~\ref{prp:simulation-FP} makes sense.  We do this by multiplying all dimensionful quantities by the appropriate powers of $x_H=\sqrt{s_H/\unitRatio_H}$, $p_H=\sqrt{s_H\unitRatio_H}$, and $\tH$ (where $\tH$, $\unitRatio_H$, and $s_H$ were defined for separable Hamiltonians in Theorem~\ref{thm:mainResultSeparable}). 
	We are not so concerned with the precise dependence on these parameters because what is interesting in Proposition~\ref{prp:simulation-FP} is the polynomial dependence in the dimension and diffusion strength -- that is, Proposition~\ref{prp:simulation-FP} is a complexity-theoretic result concerned with asymptotics at large $d$ and small $\Dz$. With these changes, \eqref{eq:sim-complexity-fp} is transformed according to 
	\begin{align}\label{eq:dimensionless-transformation}
		T\to T/\tH, 
		\Dmin\to\Dz, \Dmax\to\Dinf.
	\end{align}
	so that the time complexity to draw a single sample from $f_\auxerror$ is
	\begin{align}\label{eq:sim-complexity-fp-units}
		&\mathcal{O}\Bigg(
		\mathrm{poly}(d)
		\frac{T}{\tH}
		\Bigg(
		1+
		\te^{\frac12} \left(
		\frac{1}{\Dz^{1/4}} +
		\frac{1+\Dinf^{1/2}}{\Dz^{1/2}}
		\right) +
		\te  \frac{\Dinf^{1/2}}{\Dz^{1/2}}+
		\te^{\frac32} \frac{\Dinf+\Dinf^{3/2}}{\Dz}  +
		\te^{2} \frac{\Dinf^2}{\Dz} 
		\Bigg) \Bigg),  
	\end{align}
	where $\Dz,\Dinf = \lambda_{\mathrm{min,max}}(\D_H^{-\frac12} D \D_H^{-\frac12})$ and
	where $\te := (T/\tH)/(\error-rT)$.

	We can thus obtain Monte Carlo estimates
	of the classical expectation values 
	$\langle A\rangle_{f(T)}$ 
	using the mean of $M$ samples
	\begin{equation}
		\bar{A}_M := \frac{1}{M} \sum_{j=1}^M A(\alpha_j).
	\end{equation}
	Here, the $\alpha_j$ are sampled from the distribution $f_{\auxerror}$ that approximates $f(T)$ in the sense that $\|f_{\auxerror}-f(T)\|_{L^1}\le 2\delta$.  The $\bar{A}_M$ estimator is unbiased by construction:
	\begin{align}
		\Expec \bar{A}_M = \int A(\alpha) f_{\auxerror}(\alpha)\,\mathrm{d}\alpha = \bar{A}_\infty := \lim_{M\to\infty} \bar{A}_M,
	\end{align}
	where $\Expec$ denotes the expectation value with respect to the random samples from $f_{\auxerror}$. The error persisting in the limit of infinite samples is due to the difference between these distributions:
	\begin{align}
		\label{eq:mean-diff}
		|\bar{A}_\infty - \langle A \rangle_T |
		=\left| \int A(\alpha) (f_{\auxerror}(\alpha)-f(T,\alpha))\,\mathrm{d}\alpha\right|
		\leq \|f_{\auxerror} - f(T)\|_{L^1} \|A\|_{L^\infty} 
		= 2\auxerror \|A\|_{L^\infty} ,
	\end{align}
	The variance of the estimator $\bar{A}_M$ satisfies 
	\begin{align}
		\Expec \left[(\bar{A}_M - \bar{A}_\infty)^2 \right]
		= \frac{1}{M} \int
		(A(\alpha) - \bar{A}_\infty)^2 
		f_{\auxerror}(\alpha) \mathrm{d}\alpha
		\leq \frac{1}{M} \|A\|_{L^\infty}^2.
	\end{align}
	Using~\eqref{eq:mean-diff} and Chebyshev's inequality we obtain
	\begin{align}\label{eq:cheb}
		\Prob\left(|\bar{A}_M - \langle A\rangle_{f(T)}| \geq 4\auxerror\|A\|_{L^\infty}\right)
		\leq \Prob\left(|\bar{A}_M - \bar{A}_\infty| \geq 2\auxerror\|A\|_{L^\infty}\right)
		\leq 
		\frac{1}{4\auxerror^{2} M}.
	\end{align}
	where $\Prob$ denotes the sampling probability associated with $\Expec$.
	So if $M \geq 25 \auxerror^{-2}$ and $\|A\|_{L^\infty}\leq 1$, we can obtain an estimate of the classical expectation value 
	that is accurate to within $4\auxerror$ with probability at least $99\%$. 
	
	Therefore there exists an algorithm that computes an estimate $\bar{A}_M$ such that, with probability at least $99\%$,  
	\begin{equation}
		\label{eq:barIbd}
		\left|\bar{A}_M - \langle A\rangle_{f(T)}\right|
		\leq 4\auxerror.
	\end{equation}
	Making the choice  
	\begin{align}\label{eq:auxerror}
		\auxerror = (\error-rT)/4>0
	\end{align}
	for the total variation distance, we can combine~\eqref{eq:expectation-bd} and~\eqref{eq:barIbd} to
	conclude that the estimate $\bar{A}_M$ satisfies
	\begin{equation}
		|\Tr[\hat{A}\rho(T)] - \bar{A}_M| \leq \error
	\end{equation}
	as desired. This algorithm involves drawing $\mathcal{O}(\auxerror^{-2})$ samples from $f_\auxerror$, each of which require time complexity \eqref{eq:sim-complexity-fp-units}, so the overall complexity for the estimate is \eqref{eq:sim-complexity}.

	\subsection{Proof of Theorem~\ref{thm:simulation-simplified}} \label{sec:Lindblad-simulation-proof-simplified}
	
	Theorem~\ref{thm:simulation-simplified} follows immediately from Theorem~\ref{thm:simulation} since the latter is strictly stronger, but one can alternatively modify the proof of Theorem~\ref{thm:simulation} in the previous section to avoid using our algorithm for simulating the Fokker-Planck equation (Proposition~\ref{prp:simulation-FP}). Instead, it is already known that Euler-Murayama discrete $\tau$ time-step given by 
	\begin{equation}
		\alpha^a(t+\tau) = \alpha^a(t) + \tau (\partial^a H(\alpha(t)) 
		+ \sqrt{\tau} D^{ab}(\alpha(t)) \xi_b,
	\end{equation}
	with $\xi_b$ a standard Gaussian random variable provide an efficient (polynomial) sampling algorithm for a $\auxerror$-approximate distribution $f_\auxerror(t)$, as analyzed in~\cite{mou2022improved,dalalyan2017theoretical}. The rest of the proof (starting from \eqref{eq:auxerror}) follows as before, yielding the general polynomial time complexity as claimed for Theorem~\ref{thm:simulation-simplified} (but not the specific polynomial~\eqref{eq:sim-complexity} of Theorem~\ref{thm:simulation}).
	
	Refs.~\cite{mou2022improved,dalalyan2017theoretical} do not explicitly consider the time complexity as one takes $\Dz\to 0$, but one can recover an inverse polynomial dependence on $\Dz$ of the form $\Dz^{-6}$.  We chose to use our algorithm from Proposition~\ref{prp:simulation-FP} for the proof of Theorem~\ref{thm:simulation} because it has improved dependence on the minimum diffusion $\Dz$ for small $\Dz$ (although the complexity as a function of $T$ and $\auxerror$ is not as good). In particular, in the case of isotropic diffusion ($\Dinf=\Dz$), our algorithm has $\Dz^{-1/(2p)}$ dependence, say for $p=4$, much milder than the above $\Dz^{-6}$ for small $\Dz$.

\bibliographystyle{unsrt}
\bibliography{references}

\newpage
\begin{center}{\LARGE \textbf{Supplementary Information}}\end{center}
\renewcommand{\thesection}{\Roman{section}} 
\setcounter{section}{0}
\vspace{1em}
\vspace{2em}

\section{Simulating the Fokker-Planck equation}
\label{sec:FP-simulation-proof}

The proof of Theorem~\ref{thm:simulation} relies on Proposition~\ref{prp:simulation-FP}, which refers to a sampling algorithm for solutions to the Fokker-Planck equation $\partial_t f = v^a \partial_a f + \partial_a D^{ab} \partial_b f/2$ 
with probability density $f : \mathbb{R}^{2d} \to \mathbb{R}_{\geq 0}$, using flow field $v^a$ (varying over the domain $\mathbb{R}^{2d}$) and constant diffusion matrix $D^{ab}$. 
First we will describe the algorithm and then we will prove Proposition~\ref{prp:simulation-FP}. 

In this section, we will use $\gaussdist{\alpha,\sigma}$ instead of $\tauC_{\alpha,\sigma}$ to denote the classical normal distribution centered at $\alpha$ with covariance matrix $\sigma$.  We may also work in $\mathbb{R}^n$ for odd $n$, i.e.\ one may take ``$2d$'' to be any positive integer. We keep the notation $\mathbb{R}^{2d}$ for consistency with the rest of the paper, where we work with an even-dimensional phase space.

\subsection{Sampling algorithm for simulating the Fokker-Planck equation}
\label{sec:sampling-alg}

At uniformly separated ``respawning'' time-steps $t_k = \tau k$, for some time interval $\tau$, our algorithm produces a sequence of random vectors $\alpha_k$ and random matrices $\sigma_k$. We interpret these as the centroids and covariance matrices of Gaussian measures $\mu_k:=\mathcal{N}(\alpha_k,\sigma_k)$.
We design these measures to have the property that $\Expec\,\mu_k$ is a good approximation for $f(t_k)$, where $\Expec$ is the expectation associated with the random variables $\alpha_k$ and $\sigma_k$.  

Specifically, we begin by drawing the random variable $\alpha'_0$ from $f(0)$ and then solving the local harmonic approximation discussed around Eq.~\eqref{eq:special-case-adot} for time interval $\tau$, starting from the pair $(\alpha'_0, \sigma'_0=0)$ that corresponds to the (degenerate) Gaussian with mean $\alpha'_0$ and variance $\sigma'_0=0$.  The dynamics of the local harmonic approximation are recalled to be\footnote{When we invoke this in  Theorem~\ref{thm:simulation}, our flow field will be Hamiltonian, $v^a = \partial^a H$, and the right-hand sides will coincide with \eqref{eq:special-case-adot} and \eqref{eq:special-case-f}.}
\begin{align}
	\label{eq:reuse-adot}
	\frac{\dd \alpha^a}{\dd t} 
	&=\, v^a (\alpha),
	\\
	\label{eq:reuse-sdot}
	\frac{\dd \sigma^{ab}}{\dd t}  
	&=\,\partial_c v^a (\alpha)\sigma^{cb}+\sigma^{ac}\partial_c v^b(\alpha) + \D^{ab}.
\end{align}
It suffices to know that $v^a$ and $\partial_b v^a$ are bounded and sufficiently regular as functions of $\alpha$.
We use a $p$-th order Runge-Kutta (RK) integration (the standard scheme has $p=4$) with ``flow'' time steps of size $\zeta < \tau$ to solve the differential equations \eqref{eq:reuse-adot}, \eqref{eq:reuse-sdot} to obtain the pair $(\alpha_1, \sigma_1)$ at time $\tau$. This defines $\mu_1 := \gaussdist{\alpha_1,\sigma_1}$, so that\footnote{Remember that $\Expec$ is the expectation associated with the random variables $\alpha_k$ and $\sigma_k$. It does not indicate an average over a distribution like $\mu_k$ or $f$.} $\Expec \mu_1$ approximates $f(\tau)$. We iterate this by drawing $\alpha'_k$ from $\mu_k:=\gaussdist{\alpha_k,\sigma_k}$ and integrating to obtain $\mu_{k+1} := \gaussdist{\alpha_{k+1},\sigma_{k+1}}$ and so on, ultimately producing draws from a distribution $\Expec\, \mu_n$ that will approximate $f(T)$ for $n=T/\tau$.  (For simplicity we also assume $T$ is an integer multiple of $\tau$; if not, the final time-step can be made smaller to accommodate.)

\begin{figure}
	\centering
	\includegraphics[scale=0.6]{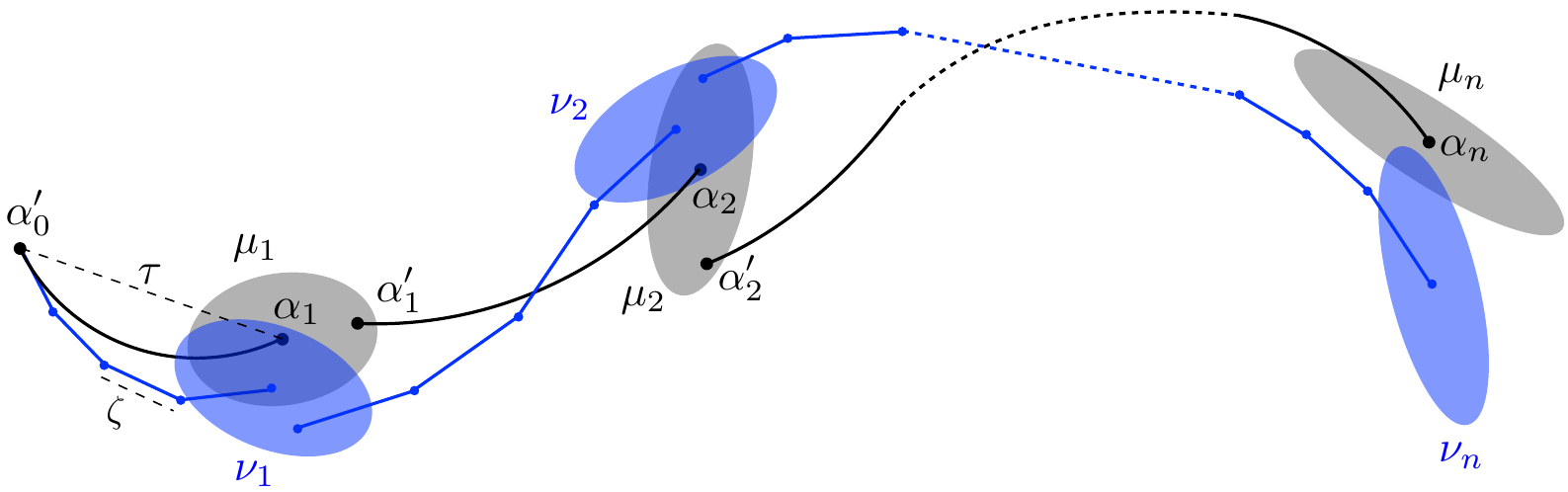}
	\caption{An illustration of the proposed approximation algorithm for sampling from $f(T)$. Starting at an initial point $\alpha'_0$ sampled from the density $f(0)$, and assuming $\sigma'_0 = 0$ we approximately solve the harmonic evolution equations~\eqref{eq:reuse-adot},~\eqref{eq:reuse-sdot} for time $\tau$ to find a new centroid $\alpha_1$ and covariance matrix $\sigma_1$, defining the Gaussian measure $\mu_1 := \mathcal{N}(\alpha_1,\sigma_1)$.  To perform this approximate evolution we propose to use a Runge-Kutta explicit integrator with time step $\zeta \leq \tau$.  We then draw a new starting points $\alpha'_1$ from $\mu_1$, integrate it forward (with $\sigma'_1=0$) to get $\alpha_{2}$ and $\sigma_{2}$, and so on. This is repeated a total of $n=T/\tau$ times, yielding a sequence random Gaussian measures $\mu_k$. The final distribution $\mu_n$, averaged over random draws, approximates $f(T)$. The exact trajectory of the Gaussian at time $\tau$ is shown in black and the approximate discretized evolution is shown in blue.  The exactly integrated Gaussian $\nu_k$ is represented by a gray ellipse, and the numerically integrated Gaussian measure $\mu_k$ by a blue ellipse. 
		So long as $\mu_k$ and $\nu_k$ have sufficient overlap the black and blue trajectories remain together.}
	\label{fig:time-stepping}
\end{figure}

This process is sketched in Figure~\ref{fig:time-stepping} and described in pseudocode in Algorithm~\ref{alg:thealg}. 
Note that the runtime of producing one sample scales as $\poly(d) \zeta^{-1}T$. 

\begin{algorithm}[H]
	\caption{Approximate sampling from $\wtild{\rho}(T)$}
	\begin{algorithmic}
		\Require $0<\zeta<\tau$, $T > 0$.
		\State $\alpha \gets $ \text{Sample}  $\rho(0)$.
		\State $t \gets 0$
		\While {$t < T$}
		\State $\alpha'\gets \alpha$, $\sigma' \gets 0$
		\State Solve harmonic evolution equations \eqref{eq:reuse-adot}, \eqref{eq:reuse-sdot} with initial 
		conditions $(\alpha',\sigma')$, for duration $\tau$,  using RK4 time-step  $\zeta$, to obtain 
		$(\alpha'(\tau),\sigma'(\tau))$.
		\State $\alpha \gets$ \text{Sample} $\mathcal{N}(\alpha'(\tau),\sigma'(\tau))$, the Gaussian distribution centered at $\alpha'(\tau)$ with 
		variance $\sigma'(\tau)$.
		\State $t \gets t+\tau$
		\EndWhile
		\State \Return $\alpha$.
	\end{algorithmic}
	\label{alg:thealg}
\end{algorithm}

\subsection{Proof of Proposition~\ref{prp:simulation-FP}}

Now we prove Proposition~\ref{prp:simulation-FP}, referring to the algorithm described in the previous subsection. We restate the theorem here for convenience.\footnote{Recall that the total variation distance between two probability distributions is $\delta(f,f')=\|f-f'\|_{L^1}/2$.} 
\prpSimulationFP*

We expect it is straightforward to extend this result to non-constant but sufficiently regular diffusion coefficients $\D$. The parameter $p$ is the order of the Runge-Kutta integrator used as a subroutine. 
This introduces a dependence on the order-$p+1$ derivatives of $H$, hidden by the $\mathcal{O}(\cdot)$ and not further analyzed.\footnote{Theorem \ref{thm:simulation} depends on $p+1$ bounded derivatives of $V$ and $\VP$ because Proposition~\ref{prp:simulation-FP} depends on $p$ derivatives of the flow field $v$.  It is also possible to use Runge-Kutta step $p=1$, although due to the  harmonic approximation invoked in the proof one must continue to require that the second derivative of $v^a$ (corresponding to the third derivative of the potential $V$) is bounded.}  
A common choice is $p=4$; larger $p$ gives better asymptotic performance but introduces more overhead and complicated dependence on higher derivatives.

We will show that for some choice of timestep sizes $\zeta$ and $\tau$ with $0<\zeta\le\tau$, the measures $\mu_n$ thus constructed satisfy
\begin{equation}
	\label{eq:f-appx-err}
	\|f(T) - \Expec\,\mu_n\|_{L^1} \leq 2\auxerror.
\end{equation}
There are two sources of error: the first is induced by the local harmonic approximation, and the second by the numerical integration of the differential equations~\eqref{eq:reuse-adot} and~\eqref{eq:reuse-sdot}. 
To separate these out, we define a new sequence of random measures $\nu_k$ which also involves a sequence of local harmonic approximations, but with exact integration of~\eqref{eq:reuse-adot} and \eqref{eq:reuse-sdot} instead of a Runge-Kutta approximation.  (We emphasize that each $\nu_{k+1}$ is computed by drawing from $\nu_k$, not $\mu_k$, and performing the exact integration.)

\subsubsection{Decomposition into two errors}

We will find it convenient to work with the classical channels -- that is, linear stochastic maps on the space of probability distributions -- that evolve the distributions. 
Define $\SFP$ be the classical channel which evolves a probability distribution according to the Fokker-Planck equation \eqref{eq:fp-sim-repeat} for some single time-step of duration $\tau$: $\SFP f(t) = f(t+\tau)$.  Define $\Snu$ be the classical channel acting on a point mass (Dirac delta function) $\gaussdist{\alpha,0}$ centered on $\alpha$ by \emph{exactly} evolving it (initially with $\sigma=0$) according to the local harmonic approximation, \eqref{eq:reuse-adot} and \eqref{eq:reuse-sdot}, also for time $\tau$; its actions on non-points masses is defined through linearity. Finally, $\Smu$ is an approximation to $\Snu$ defined by approximately integrating the local harmonic approximation \eqref{eq:reuse-adot}, \eqref{eq:reuse-sdot} with a $p$-th order Runge-Kutta method using time step $\zeta\le\tau$.  Thus, both $\Smu$ and $\Snu$ map probability distributions on phase space (viewed as linear combinations or integrals of point masses on phase space) to mixtures of Gaussian distributions. 
Furthermore, $\Expec\, \mu_n = \Smu^n \, f(0)$ approximates $\Expec\, \nu_n = \Snu^n \, f(0)$, which in turn approximates $f(T) = \SFP f(0)$.

Using the operator norm $\norm{S}_{L^1 \to L^1} := \sup_{\|f\|_{L^1}\le 1} \|Sf\|_{L^1}$ we have
\begin{align} \label{eq:S-diff}
	\norm{S - S'}_{L^1 \to L^1}
	\leq \sup_\alpha \Lonenorm{(S-S') \gaussdist{\alpha,0}}
\end{align}
for any classical channels $S, S'$.
This inequality follows from the convexity of the norm, viewing any distribution $f$ as a sum (integral) of point masses. Then, dropping the ``$L^1 \to L^1$'' subscript on the norm for clarity,
\begin{align}\begin{split}\label{eq:S-diff-iter}
		\norm{S^n - S^{\prime\, n}} &\leq \norm{(S-S') S^{n-1}  + S' (S^{n-1} - S^{\prime\, n-1})}\\
		& \leq  \norm{S - S'}  + \norm{S^{n-1} - S^{\prime\, n-1}} \leq \ldots\\
		& \leq n  \norm{S-S'}
\end{split}\end{align}
where we used the fact that a channel cannot increase $L^1$ norm.  This implies
\begin{align}
	\|f(T) - \Expec\,\mu_n\|_{L^1}
	 & \leq \|f(T) - \Expec\,\nu_n\|_{L^1} + \|\Expec\,\nu_n - \Expec\,\mu_n\|_{L^1}\\
	 & = \|(\SFP^n-\Snu^n)f(0)\|_{L^1} + \|(\Snu^n-\Smu^n)f(0)\|_{L^1} \\
	 & \leq \|\SFP^n-\Snu^n\| + \|\Snu^n-\Smu^n\| \\
	 & \leq n\left(\|\SFP-\Snu\| + \|\Snu-\Smu\|\right).
\end{align}
We will choose $\tau$ and $\zeta$ such that $\|\SFP-\Snu\|$ and $\|\Snu-\Smu\|$ are both bounded by $\auxerror/n = \tau \auxerror/T$ so that, as desired, $\|f(T) - \Expec\,\mu_n\|_{L^1}$ is bounded by $2\auxerror$. The first (harmonic approximation) error we bound essentially by applying the classical component of our main result, and the second (Runge-Kutta integration) error by a traditional error analysis in numerical ODE. 

For brevity we will not explicitly track the $\mathrm{poly}(d)$ dependence in the rest of this proof.

\subsubsection{Harmonic error}

We handle the first term by observing, using the same Duhamel argument in Step 2 of Section~\ref{sec:proof} leading to~\eqref{eq:classical-bound-special-case-with-max} and~\eqref{eq:quantum-bound-special-case-with-max}, that 
\begin{align}
	\label{eq:fp-sigma-sup}
	 \|\SFP -\Snu\| = \sup_{\|f\|_{L^1}\le 1} \|(\SFP -\Snu)f\|_{L^1}
	\leq  \tau 
	\sup_{t\leq \tau} 
	\sup_\alpha
	\Lonenorm{ \LLCRa [\tauCas]}
	\leq \mathcal{O}(\tau)\sup_{t\leq \tau} \|\exs(t)\| \|\exs(t)^{-1}\|^{\frac12} .
\end{align}
In the last inequality we used Eq.~\eqref{eq:L1_harm_error_mixed}, generalized to the case of bounded flow field $v^a$. 
Eq.~\eqref{eq:reuse-sdot} implies that 
\begin{align}
	\|\exs(t)\| & \leq \mathcal{O}( \Dmax t e^t), \label{eq:exs-bound}\\
	\|\exs(t)^{-1}\| &\leq \mathcal{O}( \Dmin^{-1} t^{-1}) \label{eq:exs-inv-bound}
\end{align}
for all $t\in[0,\tau]$, where we have assumed the flow derivative $\partial_b v^a$ is bounded by a constant.\footnote{$\|\exs(t)\|$ can blow up exponentially in time due to chaotic flow, but this will be controlled below by choosing $\tau \le 1$.}
This yields
\begin{equation} \label{eq:fTmnuT}
	 \|\SFP -\Snu\| 
	= \mathcal{O}\left(\Dmax \Dmin^{-1/2} \tau^{3/2} e^\tau\right) .
\end{equation}

\subsubsection{Runge-Kutta  integration error}

We recall the use of $p$-th order Runge-Kutta numerical integration. The ``local error,'' or the error incurred by a single time-step $\zeta$ from the initial point, goes like $\zeta^{p+1}$.  We are ultimately concerned with the ``global error,'' i.e.\ the total error in approximating the ODE evolution over time $\tau$.   In general, this error grows with $\tau$ like $\zeta^{p} e^\tau$.  (For a precise upper bound, see Theorem 3.4 of Ref.~\cite{wanner1996solving}.)  As mentioned, we will take $\tau \leq 1$ to control the exponential growth.

Consider a point mass $\gaussdist{\ina,0}$ at an arbitrary initial point  $\ina$.  Let $\exa$ and $\exs$ denote the result of exactly integrating the harmonic dynamics~\eqref{eq:reuse-adot} and~\eqref{eq:reuse-sdot}  (using zero initial variance $\ins=0$) for time $\tau$, so that $\Snu\,\gaussdist{\ina,0} = \gaussdist{\exa,\exs}$. Let $\rka$ and $\rks$ be the same for Runge-Kutta integration with time step $\zeta \le \tau$, so that $\Smu\,\gaussdist{\ina,0} = \gaussdist{\rka,\rks}$. Then the errors are bounded as\footnote{Recall that the $\mathcal{O}$ absorbs constants that depends on the order-$p$ derivatives of the flow field $v^a$, but we are retaining the dependence on $\D$.}
\begin{align}
	\label{eq:numerical-integration-error-alpha}
	\norm{\exa-\rka} 
	&= \mathcal{O}( \zeta^p \tau e^\tau), \\
	\begin{split}
		\label{eq:numerical-integration-error-sigma}
		\norm{\exs-\rks} 
		&= \mathcal{O} (\zeta^p \tau e^\tau (\Dmax+1)) 
	\end{split}
\end{align}  
To turn these estimates into $L^1$ errors for the associated Gaussian distributions we
use the following result, a slight generalization of Theorem 1.1 of Ref.~\cite{devroye2018total}.
\begin{restatable}[Distance between Gaussians]{lem}{lemSimDisGauss} \label{lemma:gaussians-distance}
	Let $\gaussdist{\beta,\Sigma}$ be the Gaussian distribution with center $\beta$ 
	and variance $\Sigma$.  Then 
	\begin{equation}
		\label{eq:gaussians-distance}
		\|\gaussdist{\beta_1,\Sigma_1} - \gaussdist{\beta_2,\Sigma_2}\|_{L^1}
		\leq \|\Sigma_1^{-1/2}(\beta_1-\beta_2)\| 
		+ \frac32 \|\Sigma_1^{-1/2}(\Sigma_2-\Sigma_1)\Sigma_1^{-1/2}\|_{\mathrm{F}}
	\end{equation}
	where the first norm is the ordinary vector 2-norm and the second is the Frobenius norm (Schatten 2-norm).
\end{restatable}
\noindent This is proved in Sec.~\ref{sec:gaussian-distance-lemma-proof}. 
From \eqref{eq:S-diff}
we obtain the following estimate, for a single time-step $\tau$:
\begin{align}
	\|\Snu-\Smu\| 
	&\leq \sup_{\ina} \Lonenorm{(\Snu-\Smu) \gaussdist{\ina,0}}\\
	&\leq \sup_{\ina} \Lonenorm{\gaussdist{\exa,\exs}-\gaussdist{\rka,\rks}}\\
	&\leq \norm{\exs^{-\frac12} (\exa-\rka)}  + \frac32 \|\exs ^{-1/2} (\rks - \exs ) \exs^{-1/2}\|_{\mathrm{F}} \\
	& \leq \norm{\exs^{-1}}^{\frac12} \norm{\exa-\rka} + \frac32 \norm{\rks - \exs} \norm{\exs^{-1}}_{\mathrm{F}} \\
	\label{eq:nu-minus-mu}
	&=\mathcal{O}( \zeta^p e^\tau (\tau^{1/2} \Dmin^{-1/2} + \Dmin^{-1}+\Dmax\Dmin^{-1})).
\end{align}
In the second line we applies \eqref{eq:S-diff}, in the third line we have applied Lemma~\ref{lemma:gaussians-distance}, and in the last line we have applied \eqref{eq:numerical-integration-error-alpha}, \eqref{eq:numerical-integration-error-sigma}, \eqref{eq:exs-bound}, and \eqref{eq:exs-inv-bound}.
The first additive term on the right-hand side is from the difference of means, and the second is from the difference of variances, corresponding to the first and second terms respectively on the right-hand side of \eqref{eq:gaussians-distance}.

\subsubsection{Sample complexity with choices for time steps $\tau$ and $\zeta$}

Now we determine choices for the time steps $\tau$ and $\zeta$ that will give us our final bound. 
The error bounds in \eqref{eq:fTmnuT} and \eqref{eq:nu-minus-mu} blow up exponentially with $\tau$, so we require $\tau \leq 1$ such that $e^\tau \le 1$. Furthermore our bound \eqref{eq:fTmnuT} on $\|\SFP-\Snu\|$ must be $\mathcal{O}(\auxerror/n) =  \mathcal{O}(\tau\auxerror/T)$, so we also require 
$\tau = \mathcal{O}((\auxerror/T)^2 \Dmax^{-2}\Dmin$.
We choose $\tau$ as large as possible consistent with these constraints, which results in some
\begin{align}\label{eq:set-tau}
	\tau 
	& = \Theta\left(\min\left\{1,(\auxerror/T)^2 \Dmax^{-2}\Dmin\right\}\right).
\end{align}
Inserting this choice of $\tau$ into \eqref{eq:nu-minus-mu} gives 
\begin{align}
	\|\Snu-\Smu\| 
	&=\mathcal{O}(\zeta^p (\min\left\{1,(\auxerror/T) \Dmax^{-1}\Dmin^{1/2}\right\} \Dmin^{-1/2} + \Dmin^{-1}+\Dmax\Dmin^{-1})).
\end{align}
For this to be upper bounded by 
$\mathcal{O}(\auxerror/n) = \mathcal{O}(\tau\auxerror/T) = \mathcal{O}\left((\auxerror/T)\min\left\{1,(\auxerror/T)^2 \Dmax^{-2}\Dmin\right\}\right)$, 
we require 
\begin{align}
	\zeta^p 
	&=\mathcal{O}\left( \frac{\auxerror}{T}\min\left\{
	\tau^{1/2}\Dmin^{1/2} , 
	\tau\Dmin,
	\tau\Dmax^{-1}\Dmin 
	\right\} \right),  \\
	&=\mathcal{O}\left( 
	\frac{\auxerror}{T}\min\left\{
	\frac{\auxerror}{T}  \Dmax^{-1}\Dmin, 
	\Dmin^{1/2} ,
	\left(\frac{\auxerror}{T}\right)^2  \Dmax^{-2}\Dmin^2,
	\left(\frac{\auxerror}{T}\right)^2  \Dmax^{-3} \Dmin^2,
	\Dmin,
	\Dmax^{-1}\Dmin
	\right\}\right).
\end{align}
It's also necessary that 
$\zeta \leq \tau = \Theta\left(\min\left\{1,(\auxerror/T)^2 \Dmax^{-2}\Dmin\right\}\right)$. 
We choose our Runge-Kutta timestep $\zeta$ to be as large as possible consistent with these constraints, resulting in some
\begin{align}\begin{split}\label{eq:zeta-choice}
	\zeta
	&=\Theta\Bigg( 
	\min\Big\{
	\left(\frac{\auxerror}{T}\right)^2  \frac{\Dmin}{\Dmax}, 
	\left(\frac{\auxerror}{T}\right)\Dmin^{1/2} ,
	\left(\frac{\auxerror}{T}\right)^3  \frac{\Dmin^2}{\Dmax^{2}},
	\left(\frac{\auxerror}{T}\right)^3 \frac{\Dmin^{2}}{\Dmax^{3}},
	\left(\frac{\auxerror}{T}\right)\Dmin,
	\left(\frac{\auxerror}{T}\right)\frac{\Dmin}{\Dmax},
	\left(\frac{\auxerror}{T}\right)^{2p} \frac{\Dmin^{p}}{\Dmax^{2p}},
	1
	\Big\}^{1/p}\Bigg),  
\end{split}\end{align}
With the choices \eqref{eq:set-tau} and \eqref{eq:zeta-choice} for $\tau$ and $\zeta$, we are ensured that $\|\SFP -\Snu\|$ and $\|\Snu -\Smu\| $ are both $\mathcal{O}\left(\auxerror/n\right)$, so that \eqref{eq:f-appx-err} can be satisfied with an appropriate choice of constant.
Using \eqref{eq:zeta-choice} for $\zeta$, the resulting time complexity $\mathcal{O}(T/\zeta)$ to compute one sample is then given by \eqref{eq:sim-complexity-fp} as desired.

\section{Proof of Lemma~\ref{lemma:gaussians-distance}}
\label{sec:gaussian-distance-lemma-proof}

We restate and prove this Lemma:
\lemSimDisGauss*
\begin{proof}
	We first we use the triangle inequality
	\begin{equation}
		\|\gaussdist{\alpha_1,\Sigma_1} - \gaussdist{\alpha_2,\Sigma_2}\|_{L^1}
		\leq 
		\|\gaussdist{\alpha_1,\Sigma_1} - \gaussdist{\alpha_2,\Sigma_1}\|_{L^1}
		+ 
		\|\gaussdist{\alpha_2,\Sigma_1} - \gaussdist{\alpha_2,\Sigma_2}\|_{L^1}.
	\end{equation}
	Then we apply translation and rescaling symmetries to observe
	\begin{align}
		\|\gaussdist{\alpha_1,\Sigma_1} - \gaussdist{\alpha_2,\Sigma_1}\|_{L^1}
		&= 
		\|\gaussdist{0,\Id} - \gaussdist{\Sigma_1^{-1/2} (\alpha_2-\alpha_1), \Id}\|_{L^1} \label{eq:tvd_diff_means}\\
		\|\gaussdist{\alpha_2,\Sigma_1} - \gaussdist{\alpha_2,\Sigma_2}\|_{L^1} 
		&= 
		\|\gaussdist{0,\Id} - \gaussdist{0, \Sigma_1^{-1/2}\Sigma_2\Sigma_1^{-1/2}}\|_{L^1}. \label{eq:tvd_diff_vars}
	\end{align}
	To further calculate \eqref{eq:tvd_diff_means}, a straightforward integral yields 
	\begin{align}
		\|\gaussdist{0,\Id} -\gaussdist{v, \Id} \|_{L^1}= 2 ( 2\Phi(\norm{v}/2)-1) \leq \frac{1}{\sqrt{2\pi}} \norm{v}
	\end{align} where $\Phi$ is the cumulative distribution function of a single-variable Gaussian with unit variance and mean zero.  To further calculate \eqref{eq:tvd_diff_vars}, we refer to Theorem 1.1 of \cite{devroye2018total}, which gives an upper bound of $\frac32 \|\Sigma_1^{-1/2}\Sigma_2\Sigma_1^{-1/2}-\Id\|_\mathrm{F}$.
	From here the lemma follows.
\end{proof}

\section{Gaussian derivatives and integrals}\label{sec:gaussian}

Here were recall basis facts about the derivatives and integrals of Gaussian functions.

\subsection{Gaussian derivatives}\label{sec:gaussian-derivatives}

The Gaussian probability distribution with mean $\alpha$ and covariance matrix $\sigma$ is  
\begin{align}
\tauCas(\alpha+\beta) &= \frac{ e^{-\beta^\tp \sigma^{-1}\beta/2}}{(2\pi)^d\sqrt{\det\sigma}} \\
&=  \frac{1}{(2\pi)^d\sqrt{\det\sigma}}\exp\left(-\frac{1}{2}\beta^a \sigma^{-1}_{ab}\beta^b\right)
\end{align}
Let us consider this a real-valued function of any vector $\beta$ and any invertible matrix $\sigma$, including non-symmetric ones, so that $\sigma_{ab}$ and $\sigma_{ba}$ are independent variables for the purposes of partial derivatives. However, at the end we will evaluate these derivatives on the subspace where $\sigma$ is symmetric. 
Recalling our notation $\partial_c = \partial/\partial\beta^c$ so $\partial_c \beta^a = \delta_c^{\pha a}$, 
we have
\begin{align}
\partial_d \left(\beta^a \sigma^{-1}_{ab}\beta^b\right) \,&=  \sigma^{-1}_{db}\beta^b + \beta^a\sigma^{-1}_{ad} 
\\
\partial_c \partial_d \left(\beta^a \sigma^{-1}_{ab}\beta^b\right) \,&=  
\sigma^{-1}_{dc} + \sigma^{-1}_{cd}.
\end{align}
We also deploy the standard \cite{petersen2012matrix}  matrix derivative identities
\begin{align}
\frac{\partial \det Z}{\partial y} &= (\det Z) \Tr\left[Z^{-1}\frac{\partial Z}{\partial y}\right],\\
\frac{\partial Z^{-1}}{\partial y} &= -Z^{-1}\frac{\partial Z}{\partial y} Z^{-1}
\end{align}
for an invertible matrix $Z$, so in particular 
\begin{align} 
\frac{\partial \det Z}{\partial Z^{ab}} &= (\det Z) Z^{-1}_{ba},\\
\frac{\partial Z^{-1}_{cd}}{\partial Z^{ab}} &= -Z^{-1}_{ca}Z^{-1}_{bd}.
\end{align}
Combining these we get
\begin{align}
\partial_a \partial_b \tauCas(\alpha+\beta) &=(\sigma^{-1}_{ac}\beta^c\sigma^{-1}_{bd}\beta^d-\sigma^{-1}_{ab})\tauCas(\alpha+\beta) \\
&= 2\frac{\partial}{\partial \sigma^{ab}} \tauCas(\alpha+\beta),
\end{align}
when evaluated for symmetric $\sigma$.  (As expected, this is singular when $\sigma$ is non-invertible.)  Quantizing both sides with the linear transformation $\WW^{-1}$ gives the corresponding quantum expression $\partial_a \partial_b \tauQas = 2\frac{\partial}{\partial \sigma_{ab}} \tauQas$.

\subsection{Gaussian integrals}\label{sec:gaussian-integrals}

Here we recall the evaluation of some Gaussian integrals, as can be done with Wick's theorem. We define the shorthand:
\begin{align}\begin{split}\label{eq:gaussian-int-cl-2}
	\langle (\beta^\tp A^{} \beta)\rangle_\sigma
	:=& \int \dd \beta \tauC_{0,\sigma}(\beta)  (\beta^\tp A^{} \beta)\\
	=&\int \dd \beta \tauCas(\alpha+\beta)  (\beta^\tp A^{} \beta) \\
	=& A^{}_{ab} \int \dd \beta \tauCas(\alpha+\beta) \beta^a\beta^b\\
	=& A^{}_{ab} \sigma^{ab}  \\
	=& \Tr[\sigma A^{}].
\end{split}\end{align}
for any positive semidefinite matrix $A$. (The covariance matrix $\sigma$ is also positive semidefinite, of course.)  Likewise, for $B$ and $C$ also positive semidefinite, we have
\begin{align}\begin{split}\label{eq:gaussian-int-cl-4}
	\langle (\beta^\tp A^{} \beta)(\beta^\tp B^{} \beta)\rangle_\sigma
	:=& \int \dd \beta \tauCas(\alpha+\beta)  (\beta^\tp A^{} \beta)(\beta^\tp B^{} \beta)\\
	= & A^{}_{ab} B^{}_{cd}  \left[\sigma^{ab}\sigma^{cd} + 2\sigma^{ad} \sigma^{bc} \right]\\
	= & \Tr[\sigma A^{}]\Tr[\sigma B^{}] + 2 \Tr[\sigma A^{} \sigma B^{}]
\end{split}\end{align}
and
\begin{align}\begin{split}\label{eq:gaussian-int-cl-6}
	\langle (\beta^\tp &A^{} \beta)(\beta^\tp B^{} \beta)(\beta^\tp C^{} \beta)\rangle_\sigma\\
	:=& \int \dd \beta \tauCas(\alpha+\beta) (\beta^\tp A^{} \beta)(\beta^\tp B^{} \beta)(\beta^\tp C^{} \beta)
	\\
	=& A^{}_{ab} B^{}_{cd} C^{}_{ef} \Big[\sigma^{ab}\sigma^{cd}\sigma^{ef} 	\\	&\qquad\qquad\qquad + 2\big(\sigma^{ab}\sigma^{cf}\sigma^{de} + \sigma^{af}\sigma^{cd}\sigma^{be} \\
	&\qquad\qquad\qquad\qquad\quad + \sigma^{ad}\sigma^{bc}\sigma^{ef}  \big) \\
	&\qquad\qquad\qquad + 4\left(\sigma^{ad}\sigma^{be}\sigma^{cf} + \sigma^{af}\sigma^{bc}\sigma^{de}  \right) \Big]\\
	=& \Tr[\sigma A^{}] \Tr[\sigma B^{}] \Tr[\sigma C^{}] + 2\Tr[\sigma A^{}] \Tr[\sigma B^{} \sigma C^{}] \\
	&\qquad +2 \Tr[\sigma B^{}] \Tr[\sigma C^{} \sigma A^{}]  + 2 \Tr[\sigma C^{}] \Tr[\sigma B^{} \sigma A^{}]
	\\
	&\qquad + 8 \Tr[\sigma A^{} \sigma B^{} \sigma C^{}] .
\end{split}\end{align}
	
\bibliographystyle{unsrt}
\bibliography{references}

\end{document}